\documentclass{elsarticle}

\usepackage{amsmath}
\usepackage{amsfonts}
\usepackage{amssymb}
\usepackage{color}
\usepackage{comment}

\usepackage{verbatim}
\usepackage{varwidth}
\usepackage{fancyvrb}

\RecustomVerbatimCommand{\VerbatimInput}{VerbatimInput}%
{fontsize=\scriptsize,
 framesep=2em, 
 rulecolor=\color{Gray},
 commandchars=\|\(\), 
 commentchar=*        
}

\begin{document}

\newcommand {\nc} {\newcommand}
  \nc {\IR} [1]{\textcolor{red}{#1}}
  \nc {\IB} [1]{\textcolor{blue}{#1}}
  \nc {\IM} [1]{\textcolor{magenta}{#1}}

\begin{frontmatter}

\title{Transfer reaction code with nonlocal interactions}

\author[MSU]{L.~J.~Titus}
\ead{titus@nscl.msu.edu}

\author[MSU]{A.~Ross}
\ead{rossal@nscl.msu.edu}

\author[MSU]{F.~M.~Nunes}
\ead{nunes@nscl.msu.edu}

\cortext[cor1]{Principal corresponding author: F.M. Nunes}
\cortext[cor2]{Corresponding author: L.J. Titus, A. Ross}

\address[MSU]{National Superconducting Cyclotron Laboratory
and Department of Physics and Astronomy,
Michigan State University, East Lansing, MI 48824, USA}

\begin{abstract}
We present a suite of codes (NLAT for nonlocal adiabatic transfer) to calculate the transfer cross section for single-nucleon transfer reactions, $(d,N)$ or $(N,d)$, including nonlocal nucleon-target interactions, within the adiabatic distorted wave approximation. For this purpose, we implement an iterative method for solving the second order nonlocal differential equation, for both scattering and bound states. The final observables that can be obtained with NLAT are differential angular distributions for the cross sections of $A(d,N)B$ or $B(N,d)A$. Details on the implementation of the T-matrix to obtain the final cross sections within the adiabatic distorted wave approximation method are also provided. 
This code is suitable to be applied for deuteron induced reactions in the range of $E_d=10-70$ MeV, and provides cross sections with $4\%$ accuracy.

\vspace{5mm}

\noindent
{\bf Program Summary}  \\
\textit{Title of package/library:} NLAT \\
\textit{Programming languages used:} Fortran~90 \\
\textit{Computers (architectures) on which the program has been tested:}\\
    Dell Poweredge R620 (Intel XEON E5-2650v2) \\
    2.9 GHz Intel Core i5 \\
\textit{Operating systems:} \\
Linux (Debian 7) \\
Mac OSX \\
\textit{RAM required to execute with typical data:} around 3~GB \\
\textit{No. of processors used:} 1 \\
\textit{CPC Library Classification:} 17.8, 17.9, 17.11, 17.16 \\
\textit{Nature of physical problem:}
    Calculates cross sections for deuteron induced single-nucleon transfer reactions using nonlocal potentials within the adiabatic distorted wave approximation. \\
\textit{Typical running time:} Less then $2$ hours.
\end{abstract}


\end{frontmatter}


\section{Introduction}

Transfer reactions are a standard probe in nuclear physics. Single-nucleon transfer in particular provides information concerning  the spin and parity of single-particle states  of the desired nucleus, as well as the probability associated with specific configurations. For this reason, these reactions are widely used in our field. Transfer reactions induced by deuterons are especially appealing because the scattering problem can be cast as a three-body problem involving  only nucleon-target interactions and the well-known NN interaction. Nevertheless, solving the three-body scattering problem $n+p+A$ exactly (e.g. \cite{Deltuva_prc2009}) is computationally intensive, and an alternative method has been proposed \cite{Johnson_npa1974}. This method, referred to as the adiabatic distorted wave approximation (ADWA), includes deuteron breakup to all orders, compares well with the exact approach \cite{Nunes_prc2011}, and has been successfully used to analyze several experiments (e.g. \cite{Jones_prc2011,Schmitt_prl2012}). 

The effective interactions between the nucleon and the composite target are a critical input to the transfer problem in ADWA. These so-called optical potentials are often extracted from elastic scattering data and made local and strongly energy dependent. However, from the microscopic point of view, it is understood that they should be nonlocal. Several studies have now demonstrated that the resulting transfer cross sections are indeed very sensitive to nonlocality \cite{Titus_prc2014,Titus_prc2015,Ross_prc2015}. As microscopic approaches to the optical potential become better suited to describe the scattering process for the isotopes of interest, it is important for the nuclear community to have access to codes that allow for the explicit inclusion of nonlocality in the optical potentials. This is the purpose of the current work. The code NLAT (NonLocal Adiabatic Transfer) provides transfer cross sections for $(d,N)$ or $(N,d)$ processes in the adiabatic distorted wave approximation. It does that through solving integro-differential wave equations through an iterative method and constructing a T-matrix with the wave functions resulting from the explicit inclusion of nonlocality.

This paper is organized in the following way. In Section \ref{sec:tmatrix} we describe the T-matrix needed to compute the transfer cross sections. In Section \ref{sec:adwa} we provide a brief description of the adiabatic distorted wave approximation along with the final expressions that were implemented in NLAT for the deuteron adiabatic distorted wave. In Section \ref{sec:WaveFunctions} we specify the wave functions that enter our transfer calculation, and in Section \ref{sec:SolvingEquation} we describe in detail the numerical method used for solving the integro-differential equation for both scattering and bound states. Computational checks on NLAT are provided in Section \ref{sec:Checks} and a guide to using the NLAT package is presented in Section \ref{sec:nlat}. Finally, we summarize and draw our conclusions in Section \ref{sec:sum}.


\section{Calculating the cross section for transfer}
\label{sec:tmatrix}

The standard way to obtain the cross section for a single-nucleon transfer reaction is through the exact T-matrix \cite{Thompson_book}. This quantity relates directly to the scattering amplitude, that when squared provides the differential cross section. We thus proceed with the exact T-matrix written in post-form \cite{Thompson_book}: the post-form is most convenient for $(d,N)$ reactions. We will also neglect the remnant term for simplicity. This term has an insignificant effect for reactions on intermediate mass and heavy nuclei (e.g. \cite{Ross_prc2015}). For the sake of clarity, we will focus all our formulation on (d,p) although a trivial reorganization of indices provides the results for the corresponding (d,n) reactions.

The exact post-form T-matrix for the $A(d,p)B$ reaction is written as:

\begin{eqnarray}\label{eq:FullTmatrix}
T_{\mu_A M_d \mu_p M_B}(\textbf{k}_f,\textbf{k}_i)=\langle \Psi_{f}^{\mu_p M_B}|V_{np}|\Psi_{i}^{\mu_A M_d}\rangle,
\end{eqnarray}

\noindent where $M_d$, $\mu_p$, $\mu_A$, and $M_B$ are the projections of the spin of the deuteron, the proton, the nucleus A, and the nucleus B, respectively. The T-matrix is related to the scattering amplitude by:

\begin{eqnarray}
f_{\mu_A M_d \mu_p M_B}(\textbf{k}_f,\textbf{k}_i)&=&-\frac{\mu_f}{2\pi\hbar^2}\sqrt{\frac{v_f}{v_i}}T_{\mu_A M_d \mu_p M_B}(\textbf{k}_f,\textbf{k}_i) ,
\end{eqnarray}

\noindent with the velocity factors given by $v_{i,f}=\hbar k_{i,f}/\mu_{i,f}$, the wave number $k_i=\sqrt{2\mu_{i,f} E_{i,f}/\hbar^2}$, $\mu_{i,f}$  the reduced mass, and $E_{i,f}$ the center of mass energy in the entrance or exit channel. The differential cross section is found by averaging the modulus squared of the scattering amplitude over initial projections, and summing over final projections,

\begin{eqnarray}
\frac{d\sigma}{d\Omega}&=&\frac{1}{\hat{J}^2_d \hat{I}^2_A}\sum_{\mu_A M_d \mu_p M_B}\left|f_{\mu_A M_d \mu_p M_B}(\textbf{k}_f,\textbf{k}_i) \right|^2 \nonumber \\
&=&\frac{k_f}{k_i}\frac{\mu_i \mu_f}{4\pi^2\hbar^4}\frac{1}{\hat{J}^2_d \hat{I}^2_A}\sum_{\mu_A M_d M_B \mu_p}\left|\langle \Psi^{\mu_p M_B}_f |V_{np}|\Psi^{\mu_A M_d}_i\rangle \right|^2,
\end{eqnarray}
\noindent where we define the quantity $\hat{X}=\sqrt{2X+1}$. 

We first focus on the initial state which describes the three-body scattering between $d+A$. As we briefly describe in Section \ref{sec:adwa}, the entrance channel wave function in the adiabatic distorted wave approximation can be expanded as:

\begin{eqnarray}\label{eq:wf5}
|\Psi_{i}^{M_d \mu_A}\rangle 
&=&\frac{4\pi}{k_i}\sum_{L_i J_{P_i}}i^{L_i}e^{i\sigma_{L_i}}\Xi_{I_A \mu_A}(\xi_A)\phi_{j_i}(r_{np})\frac{\chi_{L_i J_{p_i}}(R_{dA})}{R_{dA}}\frac{\hat{J}_{P_i}}{\hat{J}_d} \nonumber \\
&\phantom{=}& \times \  \left\{\tilde{Y}_{L_i}(\hat{k}_i)\otimes \left\{\left\{\Xi_{I_p}(\xi_p) \otimes \left\{\tilde{Y}_{\ell_i}(\hat{r}_{np}) \otimes \Xi_{I_n}(\xi_n)\right\}_{j_i} \right\}_{J_d} \right. \right. \nonumber \\
&\phantom{=}& \left. \left. \ \otimes \tilde{Y}_{L_i}(\hat{R}_{dA})\right\}_{J_{P_i}}\right\}_{J_d M_d} \nonumber \\
&=&\Xi_{I_A \mu_A}(\xi_A)\phi_{j_i}(r_{np})\chi_i^{(+)}(\textbf{k}_i,\textbf{r}_{np},\textbf{R}_{dA},\xi_p,\xi_n),
\end{eqnarray}

\noindent where $\Xi_{I_p}(\xi_p)$, $\Xi_{I_n}(\xi_n)$ and $\Xi_{I_A}(\xi_A)$ are the spin functions for the proton, neutron, and target, respectively, each with projections $\mu_p$, $\mu_n$, and $\mu_A$. $\tilde{Y}_{\ell_i}$ is the spherical harmonics for the relative motion between the neutron and proton in the deuteron, and $\tilde{Y}_{L_i}$ is the spherical harmonic for the relative motion between the deuteron and the target ($\tilde{Y}_{\ell_i}=i^{\ell_i}Y_{\ell_i}$ with $Y_{\ell_i}$ defined on p.133, Eq.(1), of \cite{Varshalovich_Book}). 

The radial bound state wave function describing the internal motion of the deuteron in the ground state, $\phi_{j_i}(r_{np})$, is the solution of the Schr\"odinger equation with potential $V_{np}$. The subscript $j_i$ results from coupling the internal orbital motion of the deuteron bound state $\ell_i$ with the spin of the neutron. $\chi_{L_i J_{p_i}}(R_{dA})$ is the radial wave function for the deuteron scattering state, with $J_{p_i}$ resulting from coupling the total angular momentum of the deuteron, $J_d=1$, to the orbital motion between the deuteron and the target, $L_i$. Section \ref{sec:adwa} describes how this wave function can be obtained in ADWA. The explicit partial wave decomposition for the incoming distorted wave is:

\begin{eqnarray}\label{eq:EntranceDistortedWave}
&\phantom{=}& \chi_i^{(+)}(\textbf{k}_i,\textbf{r}_{np},\textbf{R}_{dA},\xi_p,\xi_n)=\frac{4\pi}{k_i}\sum_{L_i J_{P_i}}i^{L_i}e^{i\sigma_{L_i}}\frac{\hat{J}_{P_i}}{\hat{J}_d}\frac{\chi_{L_i J_{p_i}}(R_{dA})}{R_{dA}} \\
&\phantom{=}& \times \ \left\{\tilde{Y}_{L_i}(\hat{k}_i)\otimes \left\{\left\{\Xi_{I_p}(\xi_p) \otimes \left\{\tilde{Y}_{\ell_i}(\hat{r}_{np}) \otimes \Xi_{I_n}(\xi_n)\right\}_{j_i} \right\}_{J_d}\otimes \tilde{Y}_{L_i}(\hat{R}_{dA})\right\}_{J_{P_i}}\right\}_{J_d M_d}. \nonumber
\end{eqnarray}

Next, we concentrate on the $p+B$ wave function in the exit channel, which is expanded as:

\begin{eqnarray}
|\Psi_{f}^{\mu_p M_B}\rangle &=&\frac{4\pi}{k_f}\left\{\Xi_{I_A}(\xi_A)\otimes \left\{ \tilde{Y}_{\ell_f}(\hat{r}_{nA})\otimes \Xi_{I_n}(\xi_n) \right\}_{j_f}\right\}_{J_B M_B}\phi_{j_f}(r_{nA})\sum_{L_f J_{P_f}}i^{L_f}e^{i\sigma_{L_f}} \nonumber \\
&\phantom{=}& \times \  \frac{\chi_{L_f J_{P_f}}(R_{pB})}{R_{pB}}\frac{\hat{J}_{P_f}}{\hat{I}_p}\left\{\tilde{Y}_{L_f}(\hat{k}_f)\otimes \left\{\Xi_{I_p}(\xi_p) \otimes \tilde{Y}_{L_f}(\hat{R}_{pB}) \right\}_{J_{P_f}} \right\}_{I_p \mu_p} \nonumber \\
&=&\left\{\Xi_{I_A}(\xi_A)\otimes \left\{ \tilde{Y}_{\ell_f}(\hat{r}_{nA})\otimes \Xi_{I_n}(\xi_n) \right\}_{j_f}\right\}_{J_B M_B}\phi_{j_f}(r_{nA})\chi_f^{(+)}(\textbf{k}_f,\textbf{R}_{pB},\xi_p). \nonumber \\
\end{eqnarray}

\noindent Here, $\ell_f$ is the orbital angular momentum between the target and the bound neutron, and $j_f$ is the quantum number resulting from coupling $\ell_f$ to the spin of the neutron,  $I_n$. The total angular momentum of the target is given by $J_B$ and results from coupling $j_f$ to $I_A$. The orbital angular momentum between the proton and the target is given by $L_f$, and the total angular momentum of the projectile, $J_{p_f}$ results from coupling $L_f$ to the spin of the proton, $I_p$. 

Note that, in Eq.(\ref{eq:FullTmatrix}), the exit channel appears as a bra:

\begin{eqnarray}\label{eq:wf11}
\langle \Psi_{f}^{\mu_p M_B}|&=&\left\{\Xi_{I_A}(\xi_A)\otimes \left\{ \tilde{Y}_{\ell_f}(\hat{r}_{nA})\otimes \Xi_{I_n}(\xi_n) \right\}_{j_f}\right\}^*_{J_B M_B}\phi_{j_f}(r_{nA})\chi_f^{(-)*}(\textbf{k}_f,\textbf{R}_{pB}), \nonumber \\
\end{eqnarray}

\noindent where the outgoing distorted wave $\chi^{(-)}(\textbf{k},\textbf{R})$ is the time reverse of $\chi^{(+)}$, so that $\chi^{(-)}(\textbf{k},\textbf{R})=\chi^{(+)}(-\textbf{k},\textbf{R})^*$. Therefore, to make this more explicit we use Eq.(2), p.141, of \cite{Varshalovich_Book},

\begin{eqnarray}
\langle \chi^{(-)}(\textbf{k},\textbf{R})|&=&\chi^{(-)*}(\textbf{k},\textbf{R}) \nonumber \\
&=&(-)^{L}\chi^{(+)}(\textbf{k},\textbf{R}),
\end{eqnarray}

\noindent where $\textbf{k}\rightarrow -\textbf{k}$ gives a factor of $(-)^{L}$ from the spherical harmonics, as seen in Eq.(2), p.141, of \cite{Varshalovich_Book}, and the two complex conjugations cancel. Therefore, the partial wave decomposition for the outgoing distorted wave is written as:

\begin{eqnarray}\label{eq:ExitDistortedWave}
&\phantom{=}& \chi_f^{(-)*}(\textbf{k}_f,\textbf{R}_{pB},\xi_p)=\frac{4\pi}{k_f\hat{I}_p}\sum_{L_f J_{P_f}}i^{-L_f}e^{i\sigma_{L_f}}\hat{J}_{P_f} \frac{\chi_{L_f J_{P_f}}(R_{pB})}{R_{pB}} \nonumber \\
&\phantom{=}& \times \ \left\{\tilde{Y}_{L_f}(\hat{k}_f)\otimes \left\{\Xi_{I_p}(\xi_p) \otimes \tilde{Y}_{L_f}(\hat{R}_{pB}) \right\}_{J_{P_f}} \right\}_{I_p \mu_p}.
\end{eqnarray}

Once the distorted waves and bound states in the initial and exit channels are determined, one can compute the T-matrix, and from it the differential cross section \cite{Thompson_book}. To begin, we insert Eq.(\ref{eq:wf5}) and Eq.(\ref{eq:wf11}) into Eq.(\ref{eq:FullTmatrix}), and make the definition,

\begin{eqnarray}
\langle \Psi^{\mu_p M_B}_f |V_{np}|\Psi^{\mu_A M_d}_i\rangle&=&\sum_{m_f}\sum_{Q M_Q}C_{I_A \mu_A j_f m_f}^{J_B M_B}C_{I_p \mu_p J_d M_d}^{Q M_Q}T_{Q M_Q m_f},
\end{eqnarray}

\noindent where $C$ are the well known Clebsch-Gordan coefficients. Then, we can conclude that:

\begin{eqnarray}
\sum_{\mu_A M_d M_B \mu_p}|\langle \Psi^{\mu_p M_B}_f |V_{np}|\Psi^{\mu_A M_d}_i\rangle|^2&=&\frac{\hat{J}_B^2}{\hat{j}_f^2}\sum_{m_f Q M_Q}T_{Q M_Q m_f}T^*_{Q M_Q m_f}.
\end{eqnarray}

\noindent After numerous algebra manipulations to simplify expressions, we arrive at the equation for the T-matrix that is implemented in the code NLAT:

\begin{eqnarray}\label{eq:tmatrix}
T_{Q M_Q m_f}&=&\frac{32\pi^3}{\sqrt{4\pi}k_ik_f}(-)^{3I_p+3I_n+j_i+J_d+3j_f}\frac{\hat{j}_f\hat{J}_{d}}{ \hat{\ell}_f}\sum_{L_i J_{P_i}} \nonumber \\
&\phantom{=}& \times \ \sum_{L_f J_{P_f}}i^{3L_i+L_f+\ell_f+\ell_i}e^{i(\sigma_{L_i}+\sigma_{L_f})}\hat{L}_i\hat{L}_f\hat{J}^2_{P_i}\hat{J}^2_{P_f} \nonumber \\
&\phantom{=}& \times \ 
\begin{Bmatrix}
I_p & J_d & j_i \\
L_f & L_i & \ell_f \\
J_{P_f} & J_{P_i} & j_f
\end{Bmatrix}\sum_{g} \hat{g}
\begin{Bmatrix}
L_f & L_i & g \\
J_{P_f} & J_{P_i} & j_f \\
I_p & J_d & Q
\end{Bmatrix}\sum_{m_g}C_{g m_g j_f m_f}^{Q M_Q}C_{L_f m_g L_i 0}^{g m_g} \nonumber \\
&\phantom{=}& \times \ Y_{L_f m_g}(\hat{k}_f)\sum_{M_K}(-)^{M_K}C_{L_f 0 L_i, -M_K}^{\ell_f,-M_K} \nonumber \\
&\phantom{=}& \times \ \int \phi_{j_f}(r_{nA})\chi_{L_f J_{P_f}}(R_{pB})V(r_{np})\phi_{j_i}(r_{np}) \nonumber \\
&\phantom{=}& \times \ \chi_{L_i J_{p_i}}(R_{dA}) \frac{R_{pB}r^2_{nA}}{R_{dA}}Y_{L_i, -M_K}(\hat{R}_{dA})Y_{\ell_f M_K}(\hat{r}_{nA})\sin\theta dR_{pB}dr_{nA}d\theta. \nonumber \\
\end{eqnarray}

The observable we compute is the differential cross section, which is obtained from Eq.(\ref{eq:tmatrix}) by:

\begin{eqnarray}\label{eq:DiffCS}
\left.\frac{d\sigma}{d\Omega}\right|_{(d,N)}&=&\frac{k_f}{k_i}\frac{\mu_i \mu_f}{4\pi^2\hbar^4}\frac{\hat{J}_B^2}{\hat{J}^2_d \hat{J}^2_A \hat{j}_f^2}\sum_{m_f Q M_Q}T_{Q M_Q m_f}T^*_{Q M_Q m_f}.
\end{eqnarray}

\noindent More details on the derivation of the final T-matrix expression can be found in \cite{thesis}. 

Eq.(\ref{eq:DiffCS}) is valid for $A(d,p)B$ reactions. The cross section for the $B(p,d)A$ reaction, where the center of mass energy of the proton in the $(d,p)$ case is identical to that in the $(p,d)$ case, is related by detailed balance:

\begin{eqnarray}
\left.\frac{d\sigma}{d\Omega}\right|_{(p,d)}=\frac{k_d^2\hat{J}_d^2\hat{I}_A^2}{k_p^2\hat{I}_p^2\hat{J}_B^2}\left.\frac{d\sigma}{d\Omega}\right|_{(d,p)}.
\end{eqnarray}

\noindent This is precisely how NLAT calculates the transfer cross sections for $(p,d)$ reactions. As was stated before, a trivial reorganization of indices provides the results for the corresponding $A(d,n)C$ reaction, with detailed balance being used to calculate the cross section for $C(n,d)A$. 

We now turn to the form in which we obtain $\phi_{j_i}(r_{np})$ and $\chi_{L_i J_{p_i}}(R_{dA})$. For this purpose, it is necessary to briefly describe the adiabatic distorted wave approximation method.


\section{Nonlocal adiabatic equation for the deuteron}
\label{sec:adwa}

The formalism for the extension of the ADWA to nonlocal interactions was presented in \cite{Titus_prc2015}. Here we summarize the theory and present details necessary for the implementation. The initial state in the exact T-matrix Eq.(\ref{eq:FullTmatrix}) is the solution of the three-body $n+p+A$ scattering problem. 
We thus begin with the corresponding three-body Schr\"odinger Equation:

\begin{equation}
\left[\hat{T}_R + T_r +V_{np}(r)+\hat{U}_{nA} + \hat{U}_{pA}-E\right]\Psi_d(\textbf{r},\textbf{R})=0,
\label{eq:adwa-3b}
\end{equation}
where $V_{np}$ is the interaction that binds the deuteron, and $U_{nA},U_{pA}$ are the effective optical potentials between the nucleons and the target.
The important realization made in \cite{Johnson_npa1974} was that when using the T-matrix of Eq.(\ref{eq:FullTmatrix}) to calculate the transfer, $\Psi_d$ is only needed in the range of $V_{np}$. For this reason,  the  adiabatic distorted wave approximation method \cite{Johnson_npa1974}  expands the wave function using Weinberg states, a basis which is only complete in the range of $V_{np}$:

\begin{eqnarray}
\Psi_d(\textbf{r},\textbf{R})=\sum_{i=0}^{\infty}\Phi_i(\textbf{r})X_i(\textbf{R}).
\end{eqnarray}
Note that in this Section, for simplicity, we use $\textbf{r}=\textbf{r}_{np}$ and $\textbf{R}=\textbf{R}_{dA}$.
One then retains only the first Weinberg state in the expansion, 
\begin{eqnarray}
\Psi(\textbf{r},\textbf{R})\approx \Phi_o(\textbf{r})X_o(\textbf{R})=\Phi(\textbf{r})X(\textbf{R}).
\end{eqnarray}

\noindent Since the first Weinberg states satisfies the equation:

\begin{equation}
(\hat{T}_r+V_{np})\Phi(\textbf{r})=-\epsilon_d \Phi(\textbf{r}),
\end{equation}

\noindent with $E_d=E+\epsilon_d$, Eq.(\ref{eq:adwa-3b}) becomes

\begin{equation}
\left[\hat{T}_R -E_d\right]\Phi(\textbf{r})X(\textbf{R})=-\left[\hat{U}_{nA} + \hat{U}_{pA}\right]\Phi(\textbf{r})X(\textbf{R}).
\label{eq:adwa1}
\end{equation}

When optical potentials are local, Eq.(\ref{eq:adwa1}) is a second-order differential equation for which direct integration methods work well. However, when the optical potentials $U_{nA}$ and $U_{pA}$ are nonlocal, the equation becomes an integro-differential equation and requires other approaches. We discuss the numerical methods in Section \ref{sec:SolvingEquation}, but for now we focus on the explicit form of the r.h.s. of Eq.(\ref{eq:adwa1}).

First we consider just the neutron potential (with $\textbf{R}_{p,n}=\textbf{R}\pm \frac{\textbf{r}}{2}$ where the ``+'' sign is for the proton and the ``-'' sign is for the neutron):

\begin{eqnarray}\label{Argument}
\hat{U}_{nA}\Psi(\textbf{r},\textbf{R})&=&\int U(\textbf{R}_n,\textbf{R}'_n)\Psi(\textbf{R}'_n, \textbf{R}_p)\delta(\textbf{R}'_p-\textbf{R}_p)d\textbf{R}'_pd\textbf{R}'_n  \\
&=&8\mathcal{J}\int U\left(\textbf{R}-\frac{\textbf{r}}{2}, \textbf{R}'-\frac{\textbf{r}'}{2}\right)\Psi(\textbf{r}',\textbf{R}')\delta \left(\textbf{r}'-(\textbf{r}-2(\textbf{R}'-\textbf{R})) \right)d\textbf{r}'d\textbf{R}' ,\nonumber
\end{eqnarray}

\noindent where the Jacobian for the coordinate transformation is

\begin{eqnarray}
\mathcal{J}=
\begin{vmatrix} 
\frac{\partial \textbf{R}'_n}{\partial\textbf{R}'} & \frac{\partial \textbf{R}'_n}{\partial \textbf{r}'} \\
\frac{\partial \textbf{R}'_p}{\partial\textbf{R}'} & \frac{\partial \textbf{R}'_p}{\partial \textbf{r}'}
\end{vmatrix}=
\begin{vmatrix}
1 & -\frac{1}{2} \\
1 & \frac{1}{2} \\
\end{vmatrix}=1.
\end{eqnarray}

\noindent Inserting the Jacobian into Eq.(\ref{Argument}) we arrive at:

\begin{eqnarray}
\hat{U}_{nA}\Psi(\textbf{r},\textbf{R})&=&8\int U_{nA}\left(\textbf{R}-\frac{\textbf{r}}{2},2\textbf{R}'-\textbf{R}-\frac{\textbf{r}}{2}\right)\Psi(\textbf{r}-2(\textbf{R}'-\textbf{R}),\textbf{R}')d\textbf{R}'. \nonumber \\
\end{eqnarray}
 
\noindent Generalizing for the nucleon nonlocal operator, and introducing the new variable $\textbf{s}=\textbf{R}'-\textbf{R}$, we obtain an adiabatic potential containing the summed effect of both the proton and neutron interactions:

\begin{eqnarray}
\label{eq:ad-pot}
&\phantom{=}&\hat{U}_{ad}\Phi(\textbf{r})X(\textbf{R})= \\
&\phantom{=}&8\int (U_{nA}\left(\textbf{R}_{n},\textbf{R}_{n}+2\textbf{s} \right)\Phi(\textbf{r} - 2\textbf{s})+U_{pA}\left(\textbf{R}_{p},\textbf{R}_{p}+2\textbf{s} \right)\Phi(\textbf{r} + 2\textbf{s}))X(\textbf{R}+\textbf{s})d\textbf{s}. \nonumber
\end{eqnarray}

Including explicitly all the angular dependence, the nonlocal adiabatic equation for the deuteron channel becomes \cite{Titus_prc2015}:

\begin{eqnarray}\label{eq:Sch-Eq-NL-ADWA}
&\phantom{=}& \left[\hat{T}_R -E_d\right]\Phi(\textbf{r})X(\textbf{R})=-\hat{U}_{ad}\Phi(\textbf{r})X(\textbf{R}) \nonumber \\
&=&-\sum_{\ell' L' J'_p}8\int U_{nA}\left(\textbf{R}_{n},\textbf{R}_{n}+2\textbf{s}  \right)\phi_{\ell' }(|\textbf{r}- 2\textbf{s}|)\frac{\chi_{L' J'_p}^{J_T M_T}\left(\left|\textbf{R}+\textbf{s} \right|\right)}{\left|\textbf{R}+\textbf{s} \right|} \nonumber \\
&\phantom{=}& \times \ \left\{ \left\{ \left\{\Xi_{I_d}(\xi_{np})\otimes \tilde{Y}_{\ell'}(\widehat{r - 2s})\right\}_{J_d} \otimes \tilde{Y}_{L'}(\widehat{R+s})\right\}_{J'_p} \otimes \Xi_{I_t}(\xi_t) \right\}_{J_T M_T}d\textbf{s}  \nonumber \\
& &-\sum_{\ell' L' J'_p}8\int U_{pA}\left(\textbf{R}_{p},\textbf{R}_{p}+2\textbf{s}  \right)\phi_{\ell' }(|\textbf{r}+ 2\textbf{s}|)\frac{\chi_{L' J'_p}^{J_T M_T}\left(\left|\textbf{R}+\textbf{s} \right|\right)}{\left|\textbf{R}+\textbf{s} \right|}  \nonumber \\
&\phantom{=}& \times \ \left\{ \left\{ \left\{\Xi_{I_d}(\xi_{np})\otimes \tilde{Y}_{\ell'}(\widehat{r + 2s})\right\}_{J_d} \otimes \tilde{Y}_{L'}(\widehat{R+s})\right\}_{J'_p} \otimes \Xi_{I_t}(\xi_t) \right\}_{J_T M_T}d\textbf{s}. \nonumber \\
\end{eqnarray}

\noindent where we defined the spin function for the deuteron to be:

\begin{equation}
\Xi_{I_d}(\xi_{np})=\left\{\Xi_{I_n}(\xi_n)\otimes \Xi_{I_p}(\xi_p) \right\}_{I_d}.
\end{equation}

The partial wave decomposition can be obtained by multiplying both sides of Eq.(\ref{eq:Sch-Eq-NL-ADWA}) by

\begin{eqnarray}\label{eq:partial-wave-decomp}
 \sum_{\ell}\left\{ \left\{ \left\{\Xi_{I_d}(\xi_{np})\otimes \tilde{Y}_{\ell}(\hat{r})\right\}_{J_d} \otimes \tilde{Y}_{L}(\hat{R})\right\}_{J_p} \otimes \Xi_{I_t}(\xi_t) \right\}_{J_T M_T}^*\phi_{\ell}(r)V_{np}(r),
\end{eqnarray}

\noindent and integrating over $d\textbf{r}$, $d\Omega_R$, $d\xi_{n}$, $d\xi_{p}$, and $d\xi_t$. 

After various steps of angular momentum re-coupling, assuming the deuteron bound state is $\ell=0$ only, and using the orthogonality properties of the Weinberg states, we  arrive at the final equation that is implemented in the code NLAT:

\begin{eqnarray}\label{eq:NLadiabaticEqn}
&\phantom{=}&\left[\frac{\hbar^2}{2\mu}\left(\frac{\partial^2}{\partial R^2}-\frac{L(L+1)}{R^2}\right) +E_d\right]\chi_{L J_p}^{J_T M_T}(R) = -\frac{8R \sqrt{\pi}}{\hat{L}} \int \phi_{0}(r)V_{np}(r) \nonumber \\
& &\times  (U_{nA}\left(\textbf{R}_{n},\textbf{R}_{n}+2\textbf{s}  \right) \phi_{0}(|\textbf{r}- 2\textbf{s}|)+U_{pA}\left(\textbf{R}_{p},\textbf{R}_{p}+2\textbf{s}  \right) \phi_{0}(|\textbf{r}+ 2\textbf{s}|))   \nonumber \\
&\phantom{=}& \times  Y_{L0}(\widehat{R+s}) \frac{\chi_{L J_p}^{J_T M_T}\left(\left|\textbf{R}+\textbf{s} \right|\right)}{\left|\textbf{R}+\textbf{s} \right|} d\textbf{s} r^2 dr \sin\theta_{r}d\theta_{r} .
\label{eq:ad-pw}
\end{eqnarray}

\begin{figure}[t!]
\begin{center}
\includegraphics[scale=0.25]{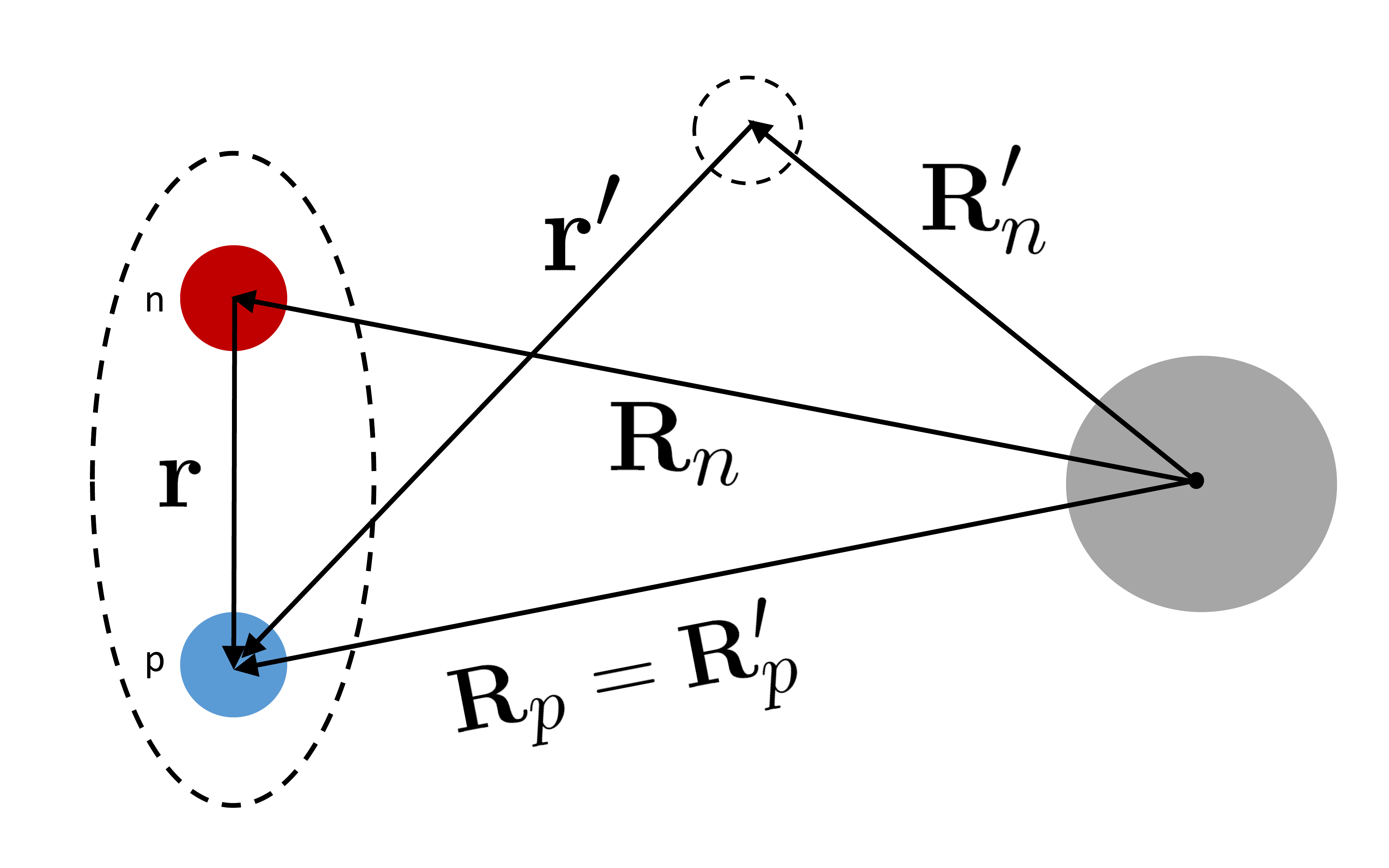}
\end{center}
\caption{The coordinates used for constructing the neutron nonlocal potential. The open dashed circle represents the neutron in a different point in space to account for nonlocality.}
\label{fig:Neutron-Nonlocal-Coordinates}
\end{figure}

The coordinates used for calculating the neutron nonlocal potential are shown in Fig.\ref{fig:Neutron-Nonlocal-Coordinates}, where the open dashed circle represents the neutron at a different point in space to account for nonlocality. While calculating the optical potential for the neutron interacting with the target, the proton remains stationary when integrating the neutron coordinate over all space. Hence, the reason for the delta function in Eq.(\ref{Argument}). 
Please see \cite{thesis} for further details.

\section{Wave functions}
\label{sec:WaveFunctions}

In order to calculate the T-matrix, Eq.(\ref{eq:FullTmatrix}), we need to define the various wave functions that go into the T-matrix, and the equations they satisfy. The entrance channel distorted wave is expanded according to Eq.(\ref{eq:EntranceDistortedWave}). The radial part of the wave function, $\chi_{L_i J_{P_i}}(R_{dA})$, satisfies the equation 


\begin{eqnarray}\label{eqn:EntranceEqn}
\left[-\frac{\hbar^2}{2\mu_i}\left(\frac{\partial^2}{\partial R_{dA}^2}-\frac{L_i(L_i+1)}{R_{dA}^2} \right)+\hat{U}_{NL}^{dA}+U_{Loc}^{dA}(R_{dA})-E_d \right]\chi_{L_i J_{P_i}}(R_{dA})=0 \nonumber \\
\label{eq:chi-in}
\end{eqnarray}

\noindent where $\hat{U}_{NL}^{dA}$ is the nonlocal part of the potential, and $U_{Loc}^{dA}(R_{dA})$ is the local part. In the {\it nonlocal ADWA}, 
$\hat{U}^{dA}_{NL}\chi_{L_iJ_{P_i}}(R_{dA})=\hat{U}_{ad}\chi_{L_iJ_{P_i}}(R_{dA})$ (the r.h.s. of Eq.(\ref{eq:NLadiabaticEqn}), with $R=R_{dA}$ and $r=r_{np}$). In practice, we only include nonlocality in the volume and surface terms of the optical potentials, $U_{nA}$ and $U_{pA}$. Therefore, there would still be a local part of the interation, $U_{Loc}^{dA}(R_{dA})$ corresponding to the spin-orbit and Coulomb terms. This is the case for the Perey-Buck \cite{Perey_np1962} and the TPM \cite{Tian_ijmpe2015} nonlocal optical potentials. 

For the {\it local ADWA}, $\hat{U}_{NL}^{dA}=0$ and $U_{Loc}^{dA}(R_{dA}) = U_{Loc}^{ad}(R_{dA}) + U_{so}(R_{dA}) + V_C(R_{dA})$, with $U_{so}$ being the sum of the neutron and proton spin-orbit potentials, $V_C$ the Coulomb potential. In this case, $U^{ad}_{Loc}(R_{dA})$ is the local adiabatic potential originally derived in \cite{Johnson_npa1974}:

\begin{equation}
U_{Loc}^{ad}(R_{dA})=-\langle \Phi_0(\textbf{r})|V_{np}(U_{nA}+U_{pA})|\Phi_0(\textbf{r})\rangle,
\end{equation}

\noindent where $\Phi_0(\textbf{r})$ is the first Weinberg state normalized such that $\langle\Phi_i|V_{np}|\Phi_{j}\rangle=-\delta_{ij}$.  

NLAT is also prepared for doing DWBA calculations. Then, the potential used to generate $\chi_{L_iJ_{P_i}}(R_{dA})$ is obtained from fits to deuteron elastic scattering, and $\hat{U}_{NL}^{dA}+U^{dA}_{Loc}(R_{dA}) \rightarrow \hat{U}_{NL}^{DWBA}+U_{Loc}^{DWBA}(R_{dA})$.  The {\it local DWBA} would be obtained by setting $\hat{U}_{NL}^{DWBA}=0$, and setting $U_{Loc}^{DWBA}(R_{dA})$ to a phenomenological deuteron optical potential, such as that from Daehnick \cite{Daehnick_prc1980}. NLAT also allows for {\it nonlocal DWBA}, a calculation in which a nonlocal deuteron optical potential is used, although currently no such parameterization is available.

As a reminder, the advantage of the ADWA is that deuteron breakup is included explicitly to all orders, and relies on much better constrained nucleon optical potentials. In the DWBA, deuteron breakup is only included implicitly through the deuteron optical potential, which is more difficult to constrain.\\

The $\ell=0$ component of the deuteron bound state satisfies the local two-body equation:
\begin{eqnarray}
\left[\hat{T}_{r_{np}} + V_{np}(r_{np})-E_{np} \right]\phi_{j_i}(r_{np})=0,
\end{eqnarray}
\noindent where $\hat{T}_{r_{np}}$ is the kinetic energy operator, $E_{np}$ is the deuteron binding energy, and the $n-p$ interaction is chosen to be a central Gaussian 
\begin{equation}
V_{np}(r_{np})=-71.85e^{-\left(\frac{r_{np}}{1.494} \right)^2}.
\end{equation}

The final state in the T-matrix Eq.(\ref{eq:FullTmatrix}) contains the final neutron bound state and the outgoing proton scattering state. The distorted wave in the exit channel is given by Eq.(\ref{eq:ExitDistortedWave}). The scattering state wave function, $\chi_{L_f J_{P_f}}(R_{pB})$, satisfies a single-channel optical model equation:
\begin{eqnarray}\label{eqn:ExitEqn}
\left[-\frac{\hbar^2}{2\mu_f}\left(\frac{\partial^2}{\partial R_{pB}^2}-\frac{L_f(L_f+1)}{R_{pB}^2} \right)+\hat{U}_{NL}^{pB}+U_{Loc}^{pB}(R_{pB})-E_p \right]\chi_{L_f J_{P_f}}(R_{pB})=0, \nonumber \\
\label{eq:chi-out}
\end{eqnarray}
\noindent with $\hat U_{NL}^{pB}$ being the nonlocal part of the potential, $U_{Loc}^{pB}(R_{pB})$ the local part, and $E_p$ the proton kinetic energy in the center of mass frame. As will be discussed shortly, the nonlocal part of the potential takes the form of Eq.(\ref{eq:kernel}), while the local part is of a Woods-Saxon form with a Coulomb potential.

The neutron bound state satisfies the equation:
\begin{eqnarray}\label{eqn:ExitBoundState}
\left[-\frac{\hbar^2}{2\mu_{nA}}\left(\frac{\partial^2}{\partial r_{nA}^2}-\frac{\ell_f(\ell_f+1)}{r_{nA}^2} \right)+\hat{V}_{NL}^{nA}+V_{Loc}^{nA}(r_{nA})-E_{nA} \right]\phi_{j_f}(r_{nA})=0 
\label{eq:phi-out}
\end{eqnarray}
\noindent where $\hat{V}_{NL}^{nA}$ is the nonlocal part of the binding potential, and $V_{Loc}^{nA}(r_{nA})$ is the local part. In NLAT, the depth of the central part of the potential needs to be adjusted to reproduce the physical neutron binding energy of the system, $E_{nA}$. 

When nonlocal potentials are used, the equations for the two scattering states and the neutron bound state are integro-differential equations. We shall next discuss how we solve them.


\section{Solving the nonlocal Sch\"odinger equation}
\label{sec:SolvingEquation}

Several methods exist for solving the nonlocal wave equation (e.g. \cite{Kim_prc1990, Rawitscher_npa2012, Michel_epja2009}). Our approach follows the iterative method proposed by Perey and Buck \cite{Perey_np1962}, and also presented in \cite{Titus_prc2014}. In this section, we will drop the local part of the nonlocal potential, containing the spin-orbit and the Coulomb, for the sake of clarity. These are included in the implementation.

The first step in solving the partial wave equation of Eqs.(\ref{eq:chi-in},\ref{eq:chi-out},\ref{eq:phi-out}) is to expand the potential  according to,
\begin{eqnarray}
U^{NL}(\textbf{R},\textbf{R}')=\sum_L\frac{2L+1}{4\pi} \frac{g_L(R,R')}{RR'}P_L(\cos \theta),
\end{eqnarray}

\noindent to obtain the kernel function $g_L(R,R')$. 

For an interaction with Gaussian nonlocality, such as the Perey and Buck \cite{Perey_np1962} interaction, there is an analytic form for the kernel function:
\begin{eqnarray}\label{eq:kernel}
g_{L}(R,R')&=& h_{L}(R,R')\;U_{WS}\left(\frac{1}{2}(R+R')\right),
\end{eqnarray}

\noindent where $U_{WS}$ is the standard local Woods-Saxon form, and 

\begin{eqnarray}\label{eq:hl}
h_{L}(R,R')&=&\frac{2i^{L}z}{\pi^{\frac{1}{2}}\beta}j_{L}(-iz)\textrm{exp}\left(-\frac{R^2+R'^2}{\beta^2}\right) \nonumber \\
	        &\approx&\frac{1}{\pi^{\frac{1}{2}}\beta}e^{-\left(\frac{R-R'}{\beta} \right)^2} \quad \textrm{for}\left|z \right| \gg 1.
\end{eqnarray}

In addition to this form, the code NLAT is prepared to read in a numerical form for $g_L(R,R')$, and thus can also be used with microscopically derived optical potentials.

\subsection{Scattering states}

In order to obtain the solution of the nonlocal equations, the iteration scheme starts with an initialization,
\begin{equation}\label{NLeqn2}
\frac{\hbar^2}{2\mu}\left[\frac{d^2}{dR^2}-\frac{L(L+1)}{R^2}\right]\chi_{n=0}(R)+[E-U_{\textrm{init}}(R)]\chi_{n=0}(R)=0,
\end{equation}

\noindent where $U_{\textrm{init}}(R)$ is a local potential. The purpose of $U_{\textrm{init}}(R)$ is to get a reasonable starting point for the iteration scheme, $\chi_o(R)$. Once we obtain $\chi_o(R)$, we proceed with solving:
\begin{eqnarray}\label{NLeqn3}
&\phantom{=}&\frac{\hbar^2}{2\mu}\left[\frac{d^2}{dr^2}-\frac{L(L+1)}{R^2} \right]\chi_n(R)+[E-U_{\textrm{init}}(R)]\chi_n(R) \nonumber \\
&=&\int g_L(R,R')\chi_{n-1}(R') dR' -U_{\textrm{init}}(R)\chi_{n-1}(R),
\end{eqnarray}

\noindent including as many iterations as necessary for convergence. While the choice for $U_{\textrm{init}}(R)$ does not affect the converged solution of the nonlocal equation, the r.h.s. of Eq.(\ref{NLeqn3}) should be small after the first iteration for such an iterative scheme to be successful. Choosing a Woods-Saxon form for $U_{\textrm{init}}(R)$ with reasonable potential parameters to approximately describe the process is sufficient for convergence. The final number of iterations depends mostly on the partial wave $L$ being solved for (small $L$ require more iterations) and the quality of $U_{\textrm{init}}(R)$. We find that when using a local equivalent potential for $U_{\textrm{init}}(R)$, convergence requires less than 10 iterations in most cases.

\subsection{Bound states}

To solve the bound state problem with a nonlocal potential we begin by solving Eq.(\ref{NLeqn2}) with a suitable $U_{\textrm{init}}(R)$ to approximate the bound state wave function for the first step on the iteration process, $\phi_o(r)$. 
We then keep track of the different normalization of the inward 
and outward wave functions resulting from the choice for the initial conditions for each wave function. 
For each iteration, we solve:

\begin{eqnarray}\label{Inward}
&\phantom{=}&\frac{\hbar^2}{2\mu}\left[\frac{d^2}{dr^2}-\frac{\ell(\ell+1)}{r^2} \right]\phi_n^{In}(r)+[E-U_{\textrm{init}}(r)]\phi_n^{In}(r) \nonumber \\
&=&\int_0^{R_{Max}} g_\ell(r,r')\phi_{n-1}^{In}(r') dr' -U_{\textrm{init}}(r)\phi_{n-1}^{In}(r),
\end{eqnarray}

\noindent and

\begin{eqnarray}\label{Outward}
&\phantom{=}&\frac{\hbar^2}{2\mu}\left[\frac{d^2}{dr^2}-\frac{\ell(\ell+1)}{r^2} \right]\phi_n^{Out}(r)+[E-U_{\textrm{init}}(r)]\phi_n^{Out}(r) \nonumber \\
&=&\int_0^{R_{Max}} g_\ell(r,r')\phi_{n-1}^{Out}(r') dr' -U_{\textrm{init}}(r)\phi_{n-1}^{Out}(r),
\end{eqnarray}

\noindent where $R_{Max}$ is some maximum radius chosen greater than the range of the nuclear interaction.  The wave function for integrating from the edge of the box inward ($\phi^{In}(r)$) is normalized to  the Whittaker function at large distances.
The wave function obtained from integrating from the origin outwards  ($\phi^{Out}(r)$) is normalized to $r^{L+1}$ close to $r=0$. These two solutions differ by a constant but both need to be tracked during the iteration process. Details can be found in \cite{thesis,Titus_prc2014}.

The condition for convergence is that the binding energy obtained from the previous iteration agrees with the 
binding energy from the current iteration within the desired level of accuracy. Due to its versatility and stability, this method provides a good option for future studies beyond Perey-Buck type potentials.

\subsection{Reading in potentials}

As was previously stated, NLAT is prepared to read in numerical forms for the kernel function $g_{L}(R,R')$ for scattering states and $g_{\ell}(r,r')$ for bound states. Let us consider bound states first. The file \textit{NLpotBound.txt} should contain the numerical form for $g_{\ell}(r,r')$ in the following format:

\begin{verbatim}
r     r'     Kernel
\end{verbatim}

\noindent This file should be placed in the same directory where NLAT will run. Have the step in $r$ and $r'$, as well as the maximum radius of $r$ and $r'$, be identical to that chosen in the input file. When making the input file, chose to include nonlocality, and then select '\textit{Read in}' when asked for what nonlocal potential to use. NLAT will still ask for the local part of the nonlocal potential, which is normally the spin-orbit and Coulomb potentials, so these parameters still need to be provided. This numerical form for $g_{\ell}(r,r')$ in \textit{NLpotBound.txt} will then be used in Eq.(\ref{Inward}) and Eq(\ref{Outward}). \\

For scattering states, the procedure is very similar. Assuming a numerical form for $g_{LJ}(R,R')$ is available for each $L$ and $J$, copy this numerical form to the file \textit{NLpotScat.txt} in the following format

\begin{verbatim}
R     R'     Real part of kernel     Imaginary part of kernel
\end{verbatim}

\noindent Notice that we allowed the kernel function to depend on both $L$ and $J$ to keep the procedure general. The file created will contain the kernel function for all partial waves. Each partial wave of the kernel functions should be saved in the same order that NLAT loops over partial waves. NLAT loops over $L$ first and then $J$. Therefore, for a spin $1/2$ projectile, the kernel functions would be ordered first the $L=0$, $J=1/2$ partial wave, second $L=1,J=1/2$, third $L=1,J=3/2$, forth $L=2,J=3/2$, and so on. Even if the nonlocal potential depends only on $L$ so that it is identical for each allowed $J$ value, still include in \textit{NLpotScat.txt} a kernel for each $LJ$ combination.  As in the bound state case, to use the numerical form for the nonlocal potential, select  '\textit{Read in}' for the nonlocal potential in the input file, and specify the local part of the nonlocal potential.


\section{Computational checks on NLAT}
\label{sec:Checks}

In this section we discuss the multiple checks  performed to ensure that NLAT produces correct and accurate results. 

\subsection{Elastic Scattering}

\begin{figure}[h!]
\begin{center}
\includegraphics[scale=0.35]{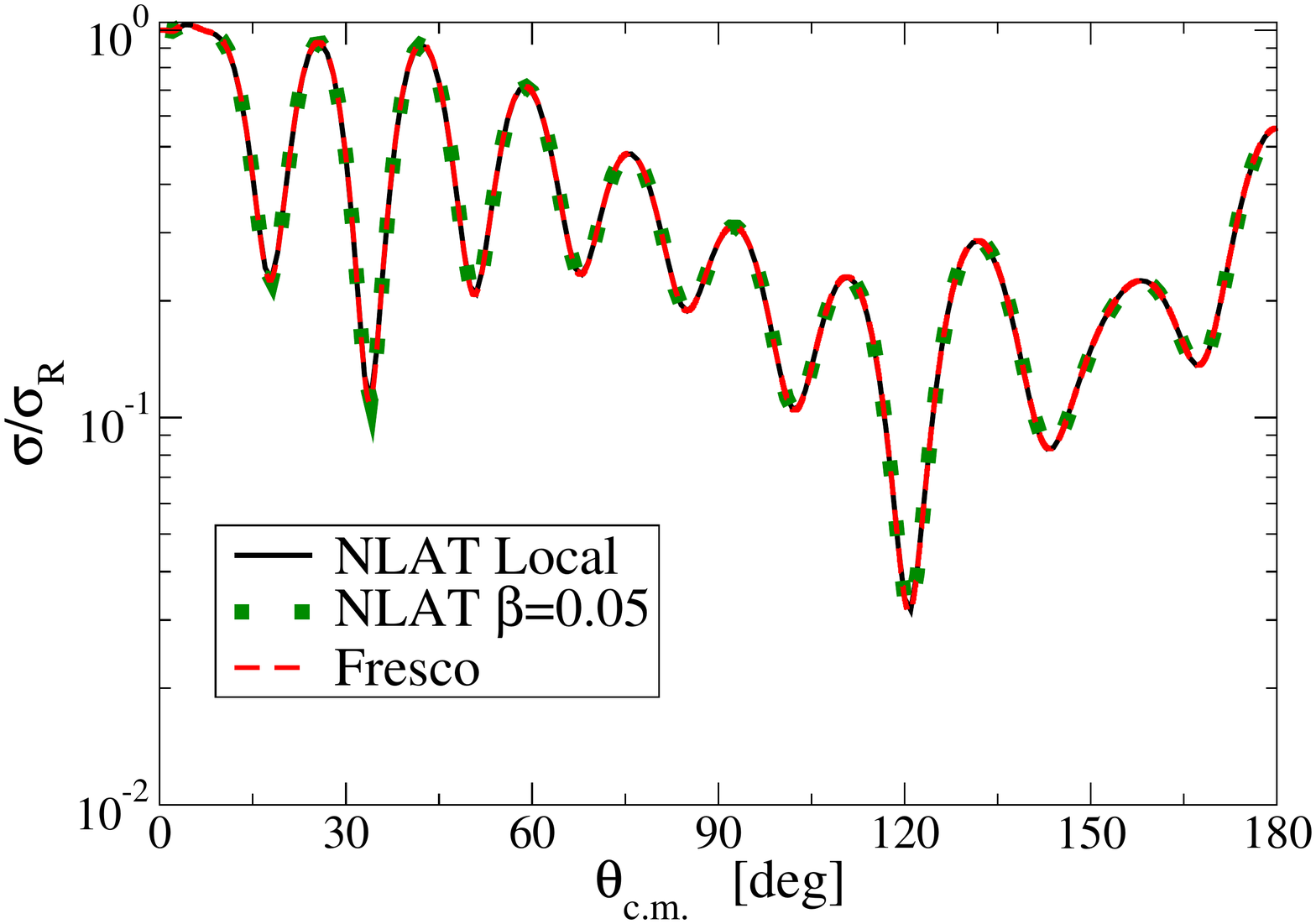}
\end{center}
\caption{Differential elastic scattering relative to Rutherford as a function of scattering angle. $^{209}$Pb$(p,p)^{209}$Pb at $E_p=50.0$ MeV: The solid line is obtained from NLAT, the dotted line is obtained from NLAT and setting $\beta=0.05$ fm, and the dashed line is from \textsc{FRESCO}. }
\label{fig:p209Pb_50-0_Local_Elastic_Check}
\end{figure}

First, we look at the local elastic scattering distribution. In Fig.\ref{fig:p209Pb_50-0_Local_Elastic_Check} we show this check for the reaction $^{209}$Pb$(p,p)^{209}$Pb at $E_p=50.0$ MeV. The solid line is a local calculation using NLAT, the dotted line is a nonlocal calculation, but with $\beta=0.05$ fm (as defined in Eq.\ref{eq:hl}) so that it approximately reduces to the local calculation, and the dashed line is the local calculation using the independent reactions code \textsc{FRESCO} \cite{fresco}. We used $\beta=0.05$ fm rather than $\beta=0$ fm since we would have numerical problems with dividing by zero if we set $\beta$ exactly equal to zero. For these calculations, we used a step size of $0.01$ fm, a maximum radius of $30$ fm, and included partial waves up to $L=20$. The numerical values for the differential cross sections differ by less than $3$\%.

\begin{figure}[h!]
\begin{center}
\includegraphics[scale=0.35]{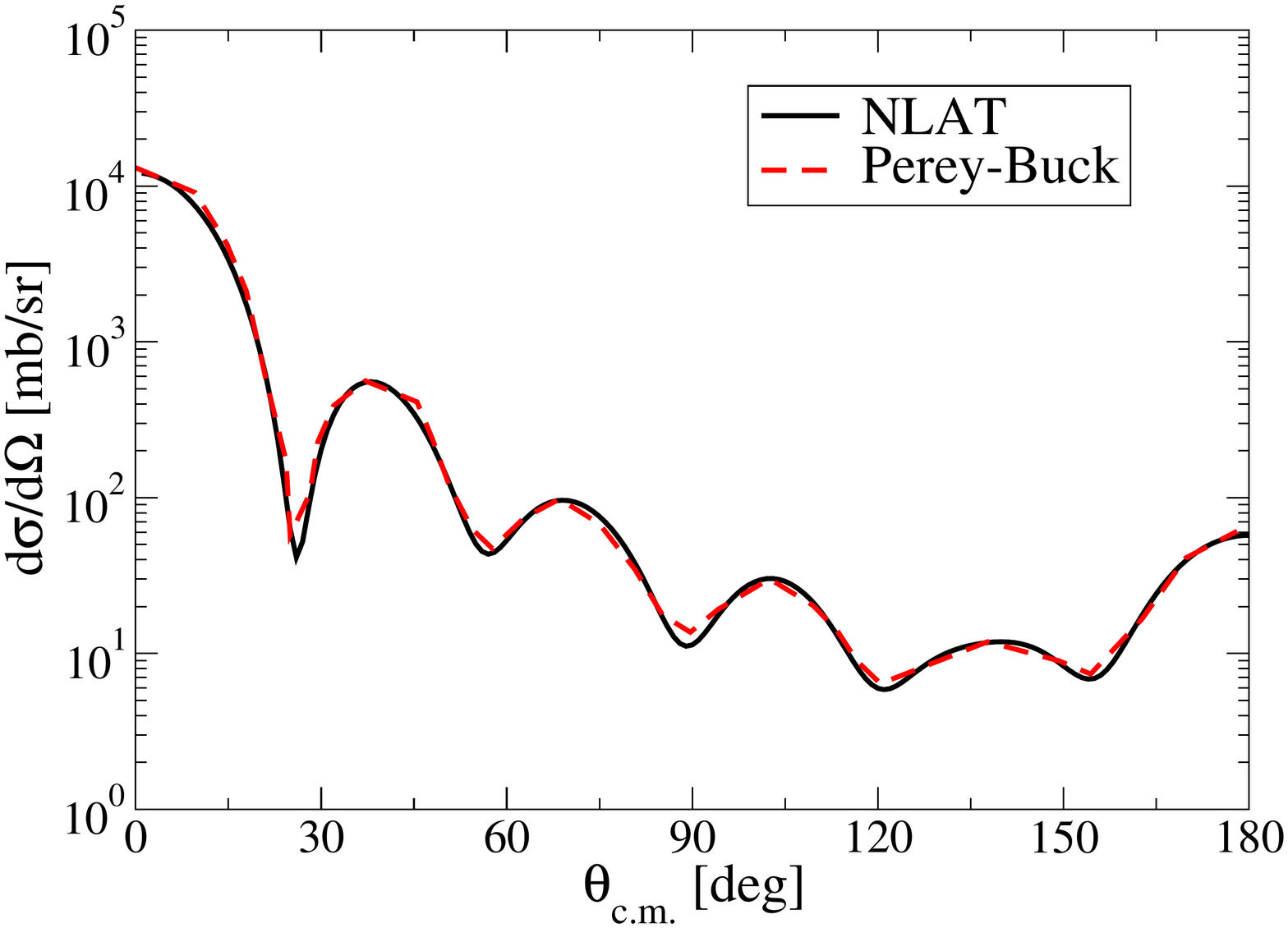}
\end{center}
\caption{Differential elastic scattering as a function of scattering angle. $^{208}$Pb$(n,n)^{208}$Pb at $E_p=14.5$ MeV: The solid line is obtained from a nonlocal calculation using NLAT, and the dashed line is the nonlocal calculation published by Perey and Buck \cite{Perey_np1962}. }
\label{fig:PereyBuck_Elastic_Check}
\end{figure}

Next, we look at the nonlocal elastic scattering distribution. In Fig.\ref{fig:PereyBuck_Elastic_Check} we present $^{208}$Pb$(n,n)^{208}$Pb at $E_n=14.5$ MeV. The solid line is a nonlocal calculation with $\beta=0.85$ fm using NLAT. The dashed line is the digitized results of the same calculation from the paper of Perey and Buck \cite{Perey_np1962}. The two calculations agree quite well, indicating that NLAT calculates elastic scattering with a nonlocal potential properly. The calculations of Perey and Buck were digitized from their paper, so any discrepancies between the results shown here and theirs is a result of errors in the digitizing process.

\subsection{Bound States}

\begin{figure}[h!]
\begin{center}
\includegraphics[scale=0.35]{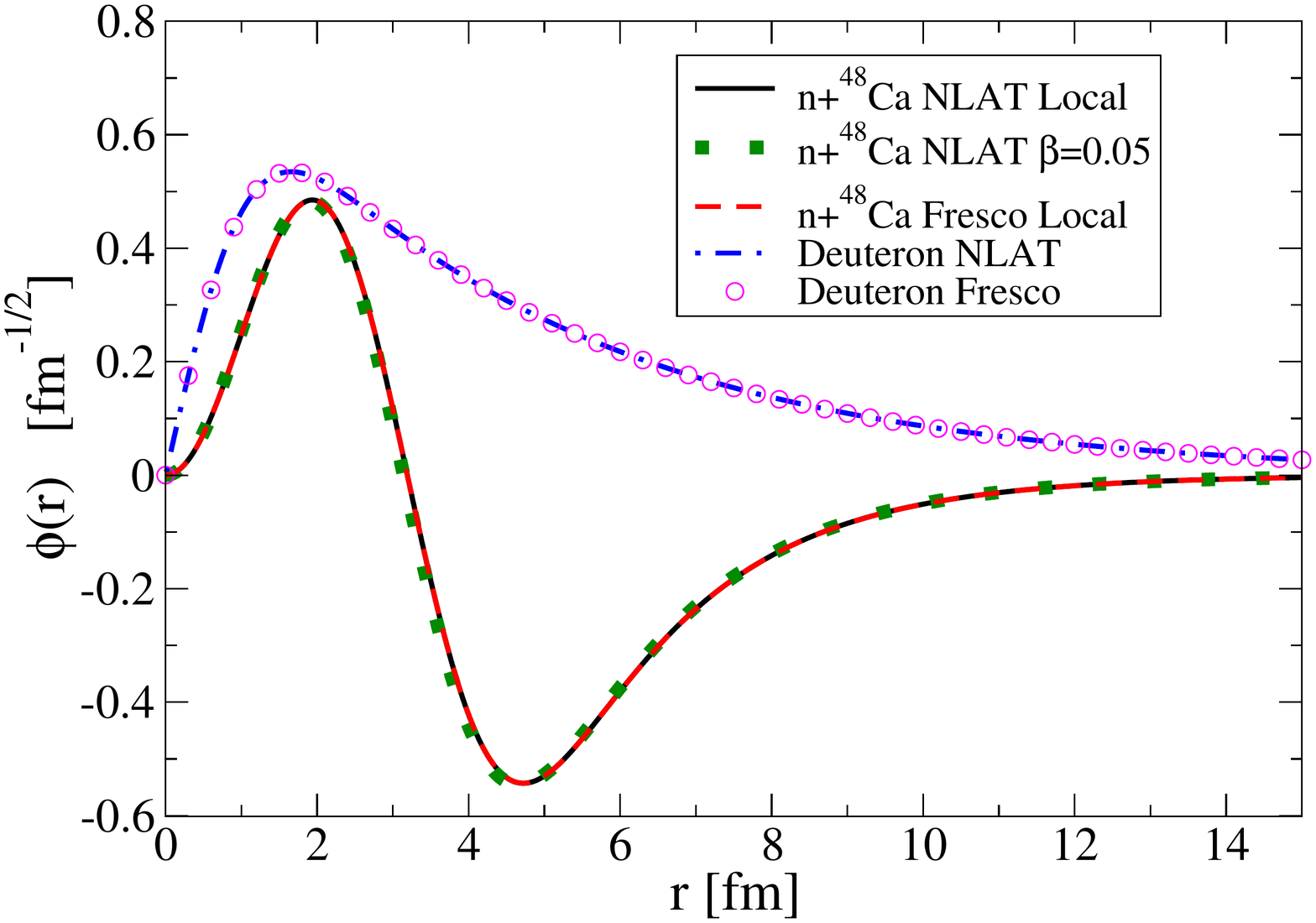}
\end{center}
\caption{$n+^{48}$Ca bound wave function, and the deuteron bound wave function. $n+^{48}$Ca: The solid line is obtained from NLAT, the dotted line is obtained from NLAT and setting $\beta=0.05$ fm, and the dashed line is from {\sc FRESCO}. Deuteron: Dot-dashed line is deuteron bound wave function obtained from NLAT, and the open circles are obtained with {\sc FRESCO}. }
\label{fig:n48Ca_Bound_Check}
\end{figure}

In order to obtain the transfer cross sections, we also need to compute the initial and final bound state wave functions. In Fig.\ref{fig:n48Ca_Bound_Check} we show the $n+^{48}$Ca bound wave function as well as the deuteron ground state wave function. For the $n+^{48}$Ca wave functions, the solid line is obtained from a local calculation with NLAT, the dotted line is a nonlocal calculation with $\beta=0.05$ fm, and the dashed line is obtained from \textsc{FRESCO}. For the deuteron bound wave function, the dot-dashed line results from a local calculation using NLAT, and the open circles are from \textsc{FRESCO}. For all calculations we used a step size of $0.01$ fm, a matching radius of $1.5$ fm, and a maximum radius of $30$ fm. With this model space the wave functions agree to at least $2$\%.

\subsection{Adiabatic Potential}
\label{Sec:Adiabatic_Potential}

\begin{figure}[h!]
\begin{center}
\includegraphics[width=\textwidth]{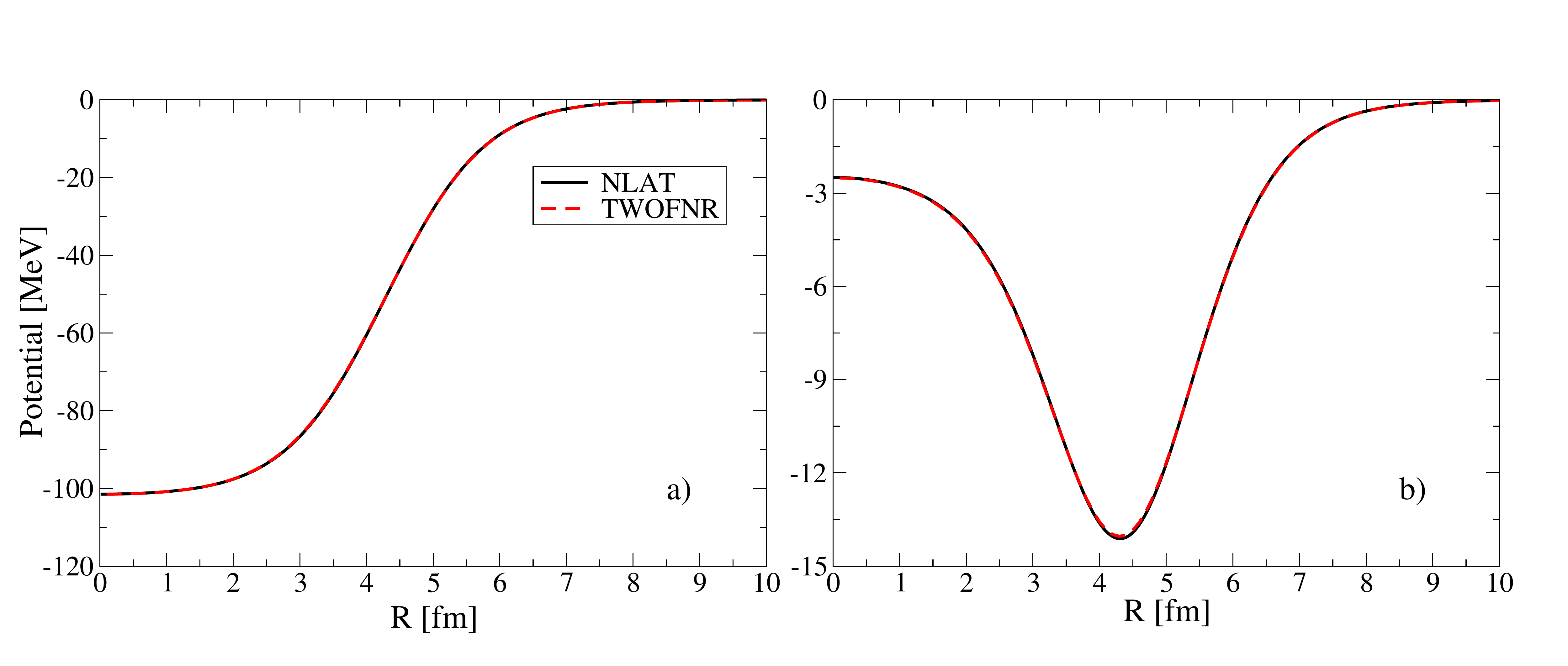}
\end{center}
\caption{The local adiabatic potential for $d+^{48}$Ca at $E_d=20$ MeV calculated with NLAT and with \textsc{TWOFNR} \cite{twofnr}. (a) Real part, (b) Imaginary part. }
\label{fig:Local_AdPot_Check}
\end{figure}

Next, we check the adiabatic potential. In Fig.\ref{fig:Local_AdPot_Check} we show the local adiabatic potential for $d+^{48}$Ca at $E_d=20$ MeV calculated with the CH89 global optical potential \cite{Varner_pr1991}. The comparison is with the code \textsc{TWOFNR} \cite{twofnr}. Panel (a) is the real part of the adiabatic potential, and (b) is the imaginary part.

Finally, we examine the resulting nonlocal source, corresponding to the r.h.s. of Eq.(\ref{eq:ad-pw}). Since we know from Fig.\ref{fig:Local_AdPot_Check} that the source agrees for $\beta \approx 0$ fm, we focus on finite values of $\beta$.  For this comparison, we used analytic expressions for the wave functions that mimicked the behavior of the numerical wave functions. For the bound wave function we used $\phi(r)=\frac{2}{r+3}e^{-0.3r}$, for the scattering wave function we used $\chi(R)=\frac{\sin(4R)}{6R}-i\frac{\sin(3R)}{5R}$,
and for the $V_{np}(r)$ potential, a central Gaussian $V_{np}(r)=-71.85e^{-\left(\frac{r}{1.494} \right)^2}$.

Mathematica \cite{Mathematica} was used to compute the source integration, and the results of this comparison are shown in Table \ref{Tab:Check_With_Mathematica}. Our results for the source are within $2$\% of the Mathematica results.

\begin{table}[h!]
\centering
\begin{tabular}{|c|r|r|r|r|}
\hline
                  &  L   & R       &  Mathematica     & NLAT           \\
\hline
$\beta=0.45$ fm   &  0   & 0.05    &  13.70+$i$16.53  &  13.71+$i$16.54   \\ 
                  &  0   & 2.00    &  1.69+$i$0.78    &  1.69+$i$0.78     \\ 
                  &  0   & 5.00    &  0.41-$i$0.29    &  0.41-$i$0.29     \\ 
                  &  1   & 0.05    &  1.67-$i$2.30    &  1.70-$i$2.33     \\ 
                  &  1   & 2.00    &  2.86+$i$1.31    &  2.86+$i$1.30     \\  
                  &  1   & 5.00    &  0.71-$i$0.50    &  0.71-$i$0.51     \\ 
                  &  5   & 0.05    &  0.00-$i$0.0002  &  -0.015+$i$0.016  \\
                  &  5   & 2.00    &  4.07+$i$1.62    &  4.08+$i$1.62     \\  
                  &  5   & 5.00    &  1.30-$i$0.91    &  1.29-$i$0.92     \\ 
\hline
$\beta=0.85$ fm   &  0   & 0.05    &  24.00-$i$23.61    & 24.01-$i$23.62  \\ 
                  &  0   & 2.00    &   2.96+$i$1.10     &  2.95+$i$1.10   \\ 
                  &  0   & 5.00    &   0.71-$i$0.33     &  0.71-$i$0.33   \\ 
                  &  1   & 0.05    &   6.52-$i$6.61     &  6.55-$i$6.64   \\ 
                  &  1   & 2.00    &   5.08+$i$1.90     &  5.08+$i$1.89   \\ 
                  &  1   & 5.00    &   1.23-$i$0.56     &  1.23-$i$0.57   \\ 
                  &  5   & 0.05    &   0.007-$i$0.0007  &  -0.02+$i$0.01  \\ 
                  &  5   & 2.00    &   8.91+$i$3.29     &  8.92+$i$3.28   \\ 
                  &  5   & 5.00    &   2.33-$i$1.06     &  2.32-$i$1.08   \\ 

\hline
\end{tabular}
\caption{The nonlocal adiabatic integral, $rhs$ of Eq.(\ref{eq:ad-pw}), calculated with Mathematica and NLAT using analytic expressions for the wave functions and potentials.}
\label{Tab:Check_With_Mathematica}
\end{table}

\subsection{Numerical Accuracy}

There were a couple of additional considerations when testing the accuracy of the method. One concerns the fact that the grid for calculating the source term was usually made coarse relative to the radial grid used to compute the wave function. Then interpolation of the source term followed. We found that while a step size of $0.01$ fm was necessary to obtain the full details of the wave function introduced in the T-matrix, good results could be obtained with a step size of $0.05$ fm in the calculation of the source term at each iteration.

\begin{figure}[h!]
\begin{center}
\includegraphics[scale=0.35]{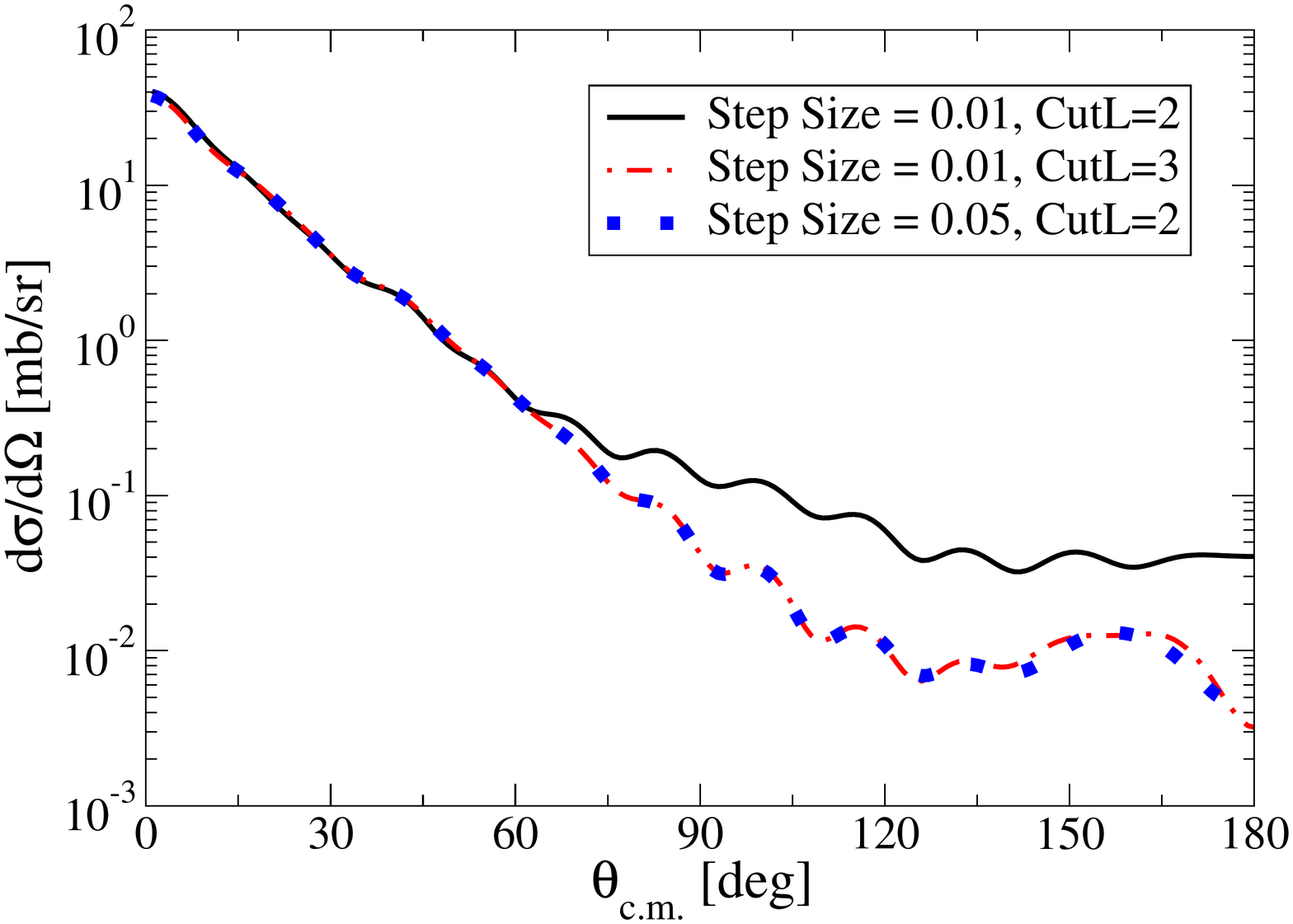}
\end{center}
\caption{Angular distributions for $^{208}$Pb$(d,p)^{209}$Pb at $E_d=50$ MeV obtained by using different step sizes and values of a cut parameter (CutL) to calculate the $rhs$ of Eq.(\ref{eq:ad-pw}). The solid line uses a step size of $0.01$ fm with CutL=2, the dashed line a step size of $0.01$ fm with CutL=3, and the square symbols use a step size of $0.05$ fm with CutL=2.}
\label{fig:CutL_Comparison}
\end{figure}

Another possible source of inaccuracy concerns the integration of the partial wave equation for high values of the angular momentum, $L$. For integrating the radial equation per partial wave, we use the Numerov method. Due to the centrifugal and Coulomb barriers, the wave functions are close to zero. In order to avoid the propagation of uncertainty, it is common to set the wave functions to zero up to a certain value of $r_i=$(StepSize)$\times$(CutL)$\times$($L$) and shift the initial radius for the integration. The variable $CutL$ in the code serves exactly this purpose. Its default value is CutL=2 but should be adjusted for small integration steps to avoid very small $r_i$.

An illustration of the interplay of the radial step used in the integration of the source term and the CutL variable is presented in Fig.\ref{fig:CutL_Comparison}. One needs to increase the CutL parameter for the $0.01$ fm calculation, from CutL=2 to CutL=3, to obtain the correct answer, in agreement with the calculation for a step of $0.05$ fm.


\section{Using the code suite `NLAT'}
\label{sec:nlat}

\subsection{General parameter Arrays}

The parameters for the reactions calculated in NLAT are contained in a series of arrays. The arrays \textbf{common} and \textbf{accuracy} are used in all types of calculations: bound, scattering, and transfer. 

\begin{itemize}
  \item{\textbf{common(:)}} \\
  Contains parameters common to all calculations
  \begin{description}
    \item{\textit{WhatCalc}: Specifies the calculation that is being performed.} \\
       1 = n+p deuteron bound state \\
       2 = N+A nucleon bound state \\
       3 = d+A deuteron scattering state \\
       4 = N+A nucleon scattering state \\
       5 = (d,N) transfer reaction \\
       6 = (N,d) transfer reaction
    \item{\textit{StepSize} [fm]: The step size that the wave functions are to be calculated in [fm]. It is highly recommended that a step size of $0.01$ fm is used.} 
    \item{\textit{Rmax} [fm]: The maximum radius the wave functions are to be calculated out to [fm]. If a transfer calculation is being performed, \textit{Rmax}, should be $30$ fm or larger.} 
    \item{\textit{ElabEntrance} [MeV]: The entrance channel projectile energy in the 
        laboratory frame for the scattering state.}
    \item{\textit{Lmax}: The maximum value of orbital angular momentum to be included in scattering calculations.}
    \item{\textit{Qvalue} [MeV]: The Q-value of the transfer reaction.}
  \end{description}
\end{itemize}

\begin{itemize}
  \item{\textbf{accuracy(:)}} \\
  Contains parameters that affect the accuracy of the calculations. The number in parentheses is the default value
  \begin{description}

    \item{\textit{MassUnit} ($931.494$  $MeV/c^2$)}: Used to calculate the reduced mass, 
      \begin{equation}
        \mu=\frac{A_1A_2}{A_1+A_2} \times MassUnit,
      \end{equation}
      with $A_1$ and $A_2$ being the mass number of the projectile (fragment) and the target (core).
    \item{\textit{npoints} ($20$): When using a Gaussian nonlocality for the potential, such as the case for the 
        integral in Eq. (\ref{NLeqn3}), the integral is solved using Gauss-Legendre quadrature using 
        \textit{npoints} number of points.}
    \item{\textit{SH\_Step} ($0.00001$): The spherical harmonics for each $\theta$ are calculated from 
        $\theta=0^\circ - 180^\circ$ in steps of \textit{SH\_Step} and stored in an array to be used in calculating the nonlocal adiabatic potential, and the T-matrix. }
    \item{\textit{Rmatchd} ($2.0 fm$): Matching radius for the deuteron bound states.}
    \item{\textit{RmatchN} ($2.5 fm$): Matching radius for the nucleon bound states.}
    \item{\textit{Estart} ($-20 \ MeV$): Energy to start scanning for the bound state.}
    \item{\textit{EnergyStep} ($0.001 \ MeV$): Step in energy when scanning for bound state.}
    \item{\textit{EnergyBackup} ($20 \ MeV$): When a given iteration for the nonlocal bound state achieves convergence, NLAT will backup by \textit{EnergyBackup} from the converged energy of the previous solution to start scanning for the solution for the next iteration.}
    \item{\textit{EdiffConvergence} ($0.001$): Percent difference at convergence between the energy of the solution of a given iteration and the energy of the previous iteration. }
    \item{\textit{convergence} ($0.001$): Percent difference of the logarithmic derivative of the bound wave function at convergence for a given iteration.}
    \item{\textit{Rstep} ($0.05 \ fm$): The step in $R$ to calculate the r.h.s. of Eq. (\ref{eq:NLadiabaticEqn}).}
    \item{\textit{Rmax} ($15 \ fm$): The maximum radius to calculate the the r.h.s. of Eq. (\ref{eq:NLadiabaticEqn}).}
    \item{\textit{lr\_max} ($4 \ fm$): The maximum radius to evaluate the integral over $r$ in the r.h.s. of Eq. (\ref{eq:NLadiabaticEqn}).}
    \item{\textit{s\_max} ($1.2 \ fm$): The maximum radius to evaluate the integral over $s$ in the r.h.s. of Eq. (\ref{eq:NLadiabaticEqn}).}
    \item{\textit{lr\_npoints} (15): The radial integral for the variable $r_{nA}$ in the nonlocal adiabatic source, 
        r.h.s. of Eq. (\ref{eq:NLadiabaticEqn}), is solved from $r_{nA}=0$ to $r_{nA}=$\textit{rnA\_Max} using 
        Gauss-Legendre quadrature with \textit{rnA\_npoints}.}
    \item{\textit{ur\_npoints} (2): The angular integral for $\theta_r$ in the nonlocal adiabatic source, 
        r.h.s. of Eq. (\ref{eq:NLadiabaticEqn}), is solved for $u_r=\cos\theta_r$ for $u_r=-1$ to $1$ using 
        Gauss-Legendre quadrature with \textit{ur\_npoints}.}
    \item{\textit{s\_npoints} (30): The radial integral for the variable $s$ in the nonlocal adiabatic source, 
        r.h.s. of Eq. (\ref{eq:NLadiabaticEqn}), is solved from $s=0$ to $s=$\textit{s\_Max} using 
        Gauss-Legendre quadrature with \textit{s\_npoints}.}
    \item{\textit{us\_npoints} (30): The angular integral for $\theta_s$ in the nonlocal adiabatic source, 
        r.h.s. of Eq. (\ref{eq:NLadiabaticEqn}), is solved for $u_s=\cos\theta_s$ for $u_s=-1$ to $1$ using 
        Gauss-Legendre quadrature with \textit{us\_npoints}.}
    \item{\textit{phi\_npoints} (6): The angular integral for $\phi_s$ in the nonlocal adiabatic source, 
        r.h.s. of Eq. (\ref{eq:NLadiabaticEqn}), is solved for $\phi_s=0-2\pi$ using 
        Gauss-Legendre quadrature with \textit{phi\_npoints}.}
    \item{\textit{RpB\_max} ($20 \ fm$): The maximum radius to evaluate the integral over $R_{pB}$ in the T-matrix equation Eq. (\ref{eq:tmatrix}). }
    \item{\textit{rnA\_max} ($20 \ fm$): The maximum radius to evaluate the integral over $r_{nA}$ in the T-matrix equation Eq. (\ref{eq:tmatrix}). }
    \item{\textit{u\_npoints} (30): The angular integral in the T-matrix, Eq. (\ref{eq:tmatrix}), is solved for 
        $u=\cos\theta$ for $u=-1$ to $1$ using Gauss-Legendre quadrature with \textit{u\_npoints}.}
    \item{\textit{RpB\_npoints} (30): The radial integral for the variable $R_{pB}$ in the T-matrix, 
        Eq. (\ref{eq:tmatrix}), is solved from $R_{pB}=0$ to $R_{pB}=$\textit{RpB\_Max} using 
        Gauss-Legendre quadrature with \textit{RpB\_npoints}.}
    \item{\textit{rNA\_npoints} (30): The radial integral for the variable $r_{nA}$ in the T-matrix, 
        Eq. (\ref{eq:tmatrix}), is solved from $r_{nA}=0$ to $r_{nA}=$\textit{rnA\_Max} using 
        Gauss-Legendre quadrature with \textit{rNA\_npoints}. }
    \item{\textit{CSstep\_transfer} ($1^\circ$): Angular step in the calculation of the transfer cross section.}
    \item{\textit{CSstep\_elastic} ($1^\circ$): Angular step in the calculation of the elastic cross section.}

  \end{description}
\end{itemize}

\subsection{Bound state parameter arrays}

The parameters for the bound states are contained in the parameter arrays \textbf{DeuteronBoundParameters} and \textbf{NucleonBoundParameters}. These will be presented generically as \textbf{BoundParameters}. The bound and scattering arrays have two indices. The first index groups parameters that describe a similar object. The second index contains the items in the lists below. For the first list below, the first item is BoundParameters(1,1), the second is BoundParameters(1,2), and so on. The same structure is retained for the scattering parameter arrays.

\begin{itemize}

  \item{\textbf{BoundParameters(1,:)}: System parameters.} 

  \begin{description}
    \item{\textit{MassFragment}: Mass number for the fragment. }
    \item{\textit{MassCore}: Mass number of the core. }
    \item{\textit{ChargeFragment}: Charge of the fragment. }
    \item{\textit{ChargeCore}: Charge of the core. }
    \item{\textit{SpinFragment}: Spin of the fragment. }
    \item{\textit{SpinCore}: Spin of the core.}
    \item{\textit{ParityFragment}: Parity of the fragment.}
    \item{\textit{ParityCore}: Parity of the core.}
    \item{\textit{L}: Orbital angular momentum of the bound state.}
    \item{\textit{jp}: Total angular momentum of the bound state. }
    \item{\textit{nn}: Number of nodes in the bound state (Minimum of 1).}
  \end{description}

  \item{\textbf{BoundParameters(2,:)}: Specifies what local potential is to be used.} 

  \begin{description}
    \item{\textit{WhatSystem}: What system is being calculated} \\
      1 = n+p deuteron bound state \\
      2 = N+A nucleon bound state \\
    \item{\textit{WhatPot}: What local binding potential is to be used} 
      \begin{description} 
        \item{Nucleon} \\
          1 = User defined local binding potential 
        \item{Deuteron} \\
          1 = Pre-defined local binding potential. 
      \end{description}
    \item{\textit{WhatPreDefPot}: What pre-defined local binding potential to use.}
      \begin{description}
        \item{Deuteron}\\
          1 = Central Gaussian. $V_{np}(r)=-71.85e^{-\left( \frac{r}{1.494} \right)^2}$
      \end{description}
  \end{description}
   
  \item{\textbf{BoundParameters(3,:)}: Specifies what nonlocal potential is to be used} 
 
  \begin{description}
    \item{\textit{NonLoc}: Determines if nonlocality is to be used.}  \\
        0 = No. Don't use nonlocality.  \\
        1 = Yes. Use nonlocality.
    \item{\textit{WhatPot}: Determines what nonlocal binding potential will to be used} 
      \begin{description} 
        \item{Nucleon} \\
          1 = User defined nonlocal binding potential  \\
          2 = Read in nonlocal binding potential
      \end{description}
  \end{description}

  \item{\textbf{BoundParameters(4,:)}: Local potential parameters.} \\
  \textit{Vv, rv, av, Vd, rvd, avd, Vso, rso, aso, rc}. \\
\\  
Contains parameters that define the local bound state potential of the Woods-Saxon type. 
  If a nonlocal calculation is to be performed, the potential defined here will take the place of 
    $U_{init}(r)$ in Eq. (\ref{Inward}) and Eq. (\ref{Outward}). 
  With these parameters, the local binding potential is defined by
    $V(r)=V_v(r)+V_d(r)+V_{so}(r)+V_C(r)$, with each term given by:

      \begin{eqnarray}\label{eq:Woods-Saxon}
        V_v(r)&=&-V_vf(r,r_v,a_v) \nonumber \\
        V_d(r)&=&4a_dV_d\frac{d}{dr}f(r,r_{vd},a_{vd}) \nonumber \\
        V_{so}(r)&=&\left(\frac{\hbar}{m_\pi c} \right)^2V_{so}\frac{1}{r}\frac{d}{dr}f(r,r_{so},a_{so})2\textbf{L}\cdot\textbf{s},
      \end{eqnarray}
 
      \noindent where 

      \begin{eqnarray}\label{eq:FormFactor}
        f(r,r_i,a_i)=\left[1+\exp\left(\frac{r-r_iA^{1/3}}{a_i} \right) \right]^{-1}.
      \end{eqnarray}
 
      \noindent The Coulomb potential is taken to be that of a homogeneous sphere of charge 

      \begin{eqnarray}\label{eq:Coulomb}
        V_C(r)=
        \begin{cases}
          \frac{Z_1 Z_2 e^2}{2}\left(3-\frac{r^2}{R_C^2} \right) & \textrm{if} \ r < R_c \\
          \frac{Z_1 Z_2 e^2}{r} & \textrm{if} \  r \geq R_c, 
        \end{cases} 
      \end{eqnarray}

      with $A$ being the mass number of the target, and the 
      Coulomb radius given by $R_C=r_cA^{1/3}$. 

  \item{\textbf{BoundParameters(5,:)}: Parameters for the local part of the nonlocal potential.} \\
  \textit{Vv, rv, av, Vd, rvd, avd, Vso, rso, aso, rc}. \\
\\  
  Contains parameters that define the local part of the nonlocal bound state potential of the Woods-Saxon type.   
  The potential defined here will take the place of $V_{Loc}^{nA}(r_{nA})$ in Eq. (\ref{eqn:ExitBoundState})
  and has the same form defined by Eq. (\ref{eq:Woods-Saxon} - \ref{eq:Coulomb}).

  \item{\textbf{BoundParameters(6,:)} Parameters for the nonlocal potential.} \\
  \textit{Vv, rv, av, Vd, rvd, avd, Vso, rso, aso, beta}. \\
  \\
  Contains parameters that define the nonlocal part of the bound state potential of the Woods-Saxon type.   
  The potential defined here specifies the $U_{WS}$ term in the kernel function of Eq. (\ref{eq:kernel}).
  This potential has the same form defined by Eq. (\ref{eq:Woods-Saxon}) and Eq. (\ref{eq:FormFactor}) 
  except with the replacement of $r \rightarrow p=\frac{r+r'}{2}$. 
  The parameter \textit{beta} holds the value of $\beta$ in Eq. (\ref{eq:hl}).

\end{itemize}

\subsection{Scattering state parameter arrays}

The parameters for the scattering states are contained in the parameter arrays \textbf{DeuteronScatParameters} and \textbf{NucleonScatParameters}. These will be presented generically as \textbf{ScatParameters}.

\begin{itemize}

  \item{\textbf{ScatParameters(1,:)}: System parameters.} 
  \begin{description}
    \item{\textit{MassProjectile}: Mass number for the projectile. }
    \item{\textit{MassTarget}: Mass number of the target. }
    \item{\textit{ChargeProjectile}: Charge of the projectile. }
    \item{\textit{ChargeTarget}: Charge of the target. }
    \item{\textit{SpinProjectile}: Spin of the projectile. }
    \item{\textit{SpinTarget}: Spin of the target.}
    \item{\textit{ParityProjectle}: Parity of the projectile.}
    \item{\textit{ParityTarget}: Parity of the target.}
  \end{description}

  \item{\textbf{ScatParameters(2,:)}: Specifies what local potential is to be used.} 
  \begin{description}
    \item{\textit{WhatSystem}: What system is being calculated} \\
      1 = Deuteron scattering state \\
      2 = Nucleon scattering state 
    \item{\textit{PotType}: What type of local scattering potential} 
      \begin{description} 
        \item{Nucleon} \\
          1 = Nucleon optical potential 
        \item{Deuteron} \\
          1 = Deuteron optical potential \\
          2 = Adiabatic potential
      \end{description}
    \item{\textit{WhatPot}: What local potential to use.} \\
      1 = User defined \\
      2 = Pre-defined
    \item{\textit{WhatPreDefPot}: What pre-defined local scattering potential to use.}
      \begin{description}
        \item{Deuteron optical potential}\\
          1 = Daehnick \cite{Daehnick_prc1980}
        \item{Nucleon optical potentials} \\
          1 = Koning-Delaroche \cite{Koning_np2003} \\
          2 = Chapel Hill \cite{Varner_pr1991}
      \end{description}
  \end{description}

  \item{\textbf{ScatParameters(3,:)}: Specifies what nonlocal potential is to be used.} 

  \begin{description}
    \item{\textit{NonLoc}: Determines if nonlocality is to be used.}  \\
          0 = No. Don't use nonlocality.  \\
          1 = Yes. Use nonlocality.
    \item{\textit{PotType}: What type of nonlocal scattering potential} 
      \begin{description} 
        \item{Nucleon} \\
          1 = Nucleon optical potential 
        \item{Deuteron} \\
          1 = Deuteron optical potential \\
          2 = Adiabatic potential
      \end{description}  
    \item{\textit{WhatPot}: What nonlocal potential to use.} \\
      1 = User defined \\
      2 = Pre-defined  \\
      3 = Read in
    \item{\textit{WhatPreDefPot}: What pre-defined local scattering potential to use.} \\
        1 = Perey-Buck \cite{Perey_np1962} \\
        2 = TPM \cite{Tian_ijmpe2015}
  \end{description}

  \item{\textbf{ScatParameters(4,:)}: Local potential parameters - two-body optical potential.} \\
  \textit{Vv, rv, av, Wv, rwv, awv, Vd, rvd, avd, Wd, rwd, awd, Vso, rso, aso, Wso, rwso, awso, rc}. \\
\\
  Contains parameters that define the local scattering state potential of the Woods-Saxon type.
  These potential parameters are used to construct the local scattering potential in the form of
    a deuteron or nucleon optical potential.
  If a nonlocal calculation is to be performed, the potential defined here will take the place of 
    $U_{init}(r)$ in Eq. (\ref{NLeqn3}). The local scattering potential is defined by
    $U(R)=U_v(R)+U_d(R)+U_{so}(R)+V_C(R)$, with each term given by:
      \begin{eqnarray}\label{eq:Woods-Saxon-Complex}
        U_v(R)&=&-V_v f(R,r_v,a_v)-iW_v f(R,r_{wv},a_{wv}) \nonumber \\
        U_d(R)&=&4a_{vd}V_d\frac{d}{dR}f(R,r_{vd},a_{vd}) +  i4a_{wd}W_d\frac{d}{dR}f(R,r_{wd},a_{wd}) \nonumber \\
        U_{so}(R)&=&\left(\frac{\hbar}{m_\pi c} \right)^2V_{so}\frac{1}{R}\frac{d}{dR}f(R,r_{so},a_{so})2\textbf{L}\cdot\textbf{s} \nonumber \\
                &\phantom{=}& + \ i\left(\frac{\hbar}{m_\pi c} \right)^2W_{so}\frac{1}{R}\frac{d}{dR}f(R,r_{wso},a_{wso})2\textbf{L}\cdot\textbf{s}
      \end{eqnarray}
      
   with the formfactors $f(R,r,a)$, and Coulomb potential $V_C(R)$ given in 
   Eq. (\ref{eq:FormFactor}) and Eq. (\ref{eq:Coulomb}) respectively. 

  \item{\textbf{ScatParameters(5,:)}: Parameters for the local part of nonlocal potential - two-body optical potential.} \\
    \textit{Vv, rv, av, Wv, rwv, awv, Vd, rvd, avd, Wd, rwd, awd, Vso, rso, aso, Wso, rwso, awso, rc}. \\
  Contains parameters that define the local part of the nonlocal scattering state potential of the Woods-Saxon type.  
  These potential parameters are used to construct the local scattering potential in the form of
    a deuteron/nucleon optical potential. 
  The potential defined here will take the place of $U_{init}$ in Eq.(37) and Eq.(38), 
  and has the same form defined by Eq. (\ref{eq:Woods-Saxon-Complex}).

  \item{\textbf{ScatParameters(6,:)} Parameters for the nonlocal part of the potential - two-body optical potential.} \\
  \textit{Vv, rv, av, Wv, rwv, awv, Vd, rvd, avd, Wd, rwd, awd, Vso, rso, aso, Wso, rwso, awso, beta}. \\
\\
  Contains parameters that define the nonlocal part of the scattering state potential of the Woods-Saxon type.  
  These potential parameters are used to construct the local scattering potential in the form of
    a deuteron/nucleon optical potential. 
  The potential defined here specifies the $U_{WS}$ term in the kernel function of Eq. (\ref{eq:kernel}).
  This potential has the same form defined by Eq. (\ref{eq:Woods-Saxon-Complex}) 
    except with the replacement of $R \rightarrow p=\frac{R+R'}{2}$. 
  The parameter \textit{beta} holds the value of $\beta$ in Eq. (\ref{eq:hl}).   

  \item{\textbf{ScatParameters(7,:)} and \textbf{ScatParameters(8,:)}: Local potential parameters - neutron or proton optical potential for the adiabatic potential.} \\
  \textit{Vv, rv, av, Wv, rwv, awv, Vd, rvd, avd, Wd, rwd, awd, Vso, rso, aso, Wso, rwso, awso, rc}. \\
\\ 
  Same as \textbf{ScatParameters(4,:)} except these hold the parameters that define the 
    local optical potential for the proton and the neutron in the deuteron at half the deuteron energy.
  \textbf{ScatParameters(7,:)} is for the neutron, and \textbf{ScatParameters(8,:)} is for the proton.

  \item{\textbf{ScatParameters(9,:)} and \textbf{ScatParameters(10,:)}: Parameters for the local part of nonlocal potential - neutron or proton optical potential for the adiabatic potential.} \\
  \textit{Vv, rv, av, Wv, rwv, awv, Vd, rvd, avd, Wd, rwd, awd, Vso, rso, aso, Wso, rwso, awso, rc}. \\
\\
  Same as \textbf{ScatParameters(5,:)} except these hold the parameters that define the 
    local part of the nonlocal potential for the proton and the neutron in the deuteron.
  \textbf{ScatParameters(9,:)} is for the neutron, and \textbf{ScatParameters(10,:)} is for the proton.

  \item{\textbf{ScatParameters(11,:)} and \textbf{ScatParameters(12,:)}: Parameters for the nonlocal part of the potential - neutron or proton optical potential for the adiabatic potential.} \\
  \textit{Vv, rv, av, Wv, rwv, awv, Vd, rvd, avd, Wd, rwd, awd, Vso, rso, aso, Wso, rwso, awso, beta}. \\
\\
  Same as \textbf{ScatParameters(6,:)} except these hold the parameters that define the 
    local part of the nonlocal potential for the proton and the neutron in the deuteron in ADWA.
  \textbf{ScatParameters(11,:)} is for the neutron, and \textbf{ScatParameters(12,:)} is for the proton.

\end{itemize}

\subsection{Making an input file}

Contained in the source code is a small program called \textit{make\_input.f90}. 
Compile this code with either the gfortran or ifort compilers. Run the executable in order to make an input file (see Section \ref{sec-compile}) 
The code will prompt the user to enter the necessary values by asking a series of questions.
The questions asked are written into the input file to the right of each value on each line, as is seen in the example input files below.
The output of the code is written into the file \textit{inputfile.in}. 
If a mistake is made while making the input, the user can back up by copying \textit{inputfile.in}
  to the file \textit{temp.in}. The user can then delete the mistakes that were made. 
  When \textit{make\_input.f90} is re-ran, it will read from the file \textit{temp.in} until there 
  are no more value to read, and then continue prompting questions.

\subsection{$^{48}$Ca$(d,p)^{49}$Ca at $E_d = 20$ MeV}

Here we will give an example of the reaction $^{48}$Ca$(d,p)^{49}$Ca at $E_d = 20$ MeV. 
The input file \textit{dp48Ca\_20-0\_NL.in} was created 

\IB{ 
\begin{Verbatim}
5                       What Calculation? 1=n+p (B), 2=N+A (B), 3=d+A (S), 4=N+A (S), 5=(d,N), 6=(N,d)
 
0.010                   Step Size
30.000                  Maximum Radius
20.000                  Elab
20                      Lmax
2.946                   Q-Value
 
1                       Pre-Defined Deuteron Binding Potential. [1]=Gaussian
 
1     48                Mass number of fragement and core (N+A BOUND STATE)
0     20                Charge of fragement and core
0.5   0.0               Spin of fragement and core
1.0   1.0               Parity of fragement and core
1     1.5               L and J of Bound State
2                       Number of nodes in bound state?
1                       What Local Potential To Use? (1=User defined)
50.301   1.250  0.650   Local Real Volume Depth, Radius, Diffuseness
0.000    0.000  0.000   Local Real Surface Depth, Radius, Diffuseness
6.000    1.250  0.650   Local Real Spin-Orbit Depth, Radius, Diffuseness
1.250                   Local Coulomb Radius
1                       Use Nonlocality? [0]=No, [1]=Yes
1                       What Nonlocal Potential? (1=User defined, 2=Read in)
0.000    0.000  0.000   Local Part of NL: Real Volume Depth, Radius, Diffuseness
0.000    0.000  0.000   Local Part of NL: Real Surface Depth, Radius, Diffuseness
6.000    1.250  0.650   Local Part of NL: Real Spin-Orbit Depth, Radius, Diffuseness
1.220                   Local Part of NL: Coulomb Radius
63.040   1.250  0.650   Nonlocal Real Volume Depth, Radius, Diffuseness
0.000    0.000  0.000   Nonlocal Real Surface Depth, Radius, Diffuseness
0.000    0.000  0.000   Nonlocal Real Spin-Orbit Depth, Radius, Diffuseness
0.850                   Beta: Range of Nonlocality
 
2     48                Mass number of projectile and target (NUCLEON SCATTERING STATE)
1     20                Charge of projectile and target
1.0   0.0               Spin of projectile and target
1.0   1.0               Parity of projectile and target
1                       What Local Potential? (1=Deuteron Optical Potential, 2=Nucleon Potentials)
1                       What Local Potential To Use? (1=User defined, 2=Pre-defined)
97.842   1.251  0.629   Local Real Volume Depth, Radius, Diffuseness
0.000    0.000  0.000   Local Imag Volume Depth, Radius, Diffuseness
0.000    0.000  0.000   Local Real Surface Depth, Radius, Diffuseness
8.697    1.236  0.440   Local Imag Surface Depth, Radius, Diffuseness
7.180    1.220  0.650   Local Real Spin-Orbit Depth, Radius, Diffuseness
0.000    0.000  0.000   Local Imag Spin-Orbit Depth, Radius, Diffuseness
1.220                   Local Coulomb Radius
1                       Use Nonlocality? [0]=No, [1]=Yes
2                       What Type of Nonlocal Potential? (1=Deuteron Potential, 2=Nucleon Potentials)
1                       What Local Nucleon Potentials To Use? (1=User defined, 2=Pre-defined)
0.000    0.000  0.000   Local Part of NL: Neutron Real Volume Depth, Radius, Diffuseness
0.000    0.000  0.000   Local Part of NL: Neutron Imag Volume Depth, Radius, Diffuseness
0.000    0.000  0.000   Local Part of NL: Neutron Real Surface Depth, Radius, Diffuseness
0.000    0.000  0.000   Local Part of NL: Neutron Imag Surface Depth, Radius, Diffuseness
7.180    1.220  0.650   Local Part of NL: Neutron Real Spin-Orbit Depth, Radius, Diffuseness
0.000    0.000  0.000   Local Part of NL: Neutron Imag Spin-Orbit Depth, Radius, Diffuseness
0.000    0.000  0.000   Local Part of NL: Proton Real Volume Depth, Radius, Diffuseness
0.000    0.000  0.000   Local Part of NL: Proton Imag Volume Depth, Radius, Diffuseness
0.000    0.000  0.000   Local Part of NL: Proton Real Surface Depth, Radius, Diffuseness
0.000    0.000  0.000   Local Part of NL: Proton Imag Surface Depth, Radius, Diffuseness
7.180    1.220  0.650   Local Part of NL: Proton Real Spin-Orbit Depth, Radius, Diffuseness
0.000    0.000  0.000   Local Part of NL: Proton Imag Spin-Orbit Depth, Radius, Diffuseness
1.220                   Local Part of NL: Proton Coulomb Radius
71.000   1.220  0.650   Nonlocal Neutron Real Volume Depth, Radius, Diffuseness
0.000    0.000  0.000   Nonlocal Neutron Imag Volume Depth, Radius, Diffuseness
0.000    0.000  0.000   Nonlocal Neutron Real Surface Depth, Radius, Diffuseness
15.000   1.220  0.470   Nonlocal Neutron Imag Surface Depth, Radius, Diffuseness
0.000    0.000  0.000   Nonlocal Neutron Real Spin-Orbit Depth, Radius, Diffuseness
0.000    0.000  0.000   Nonlocal Neutron  Imag Spin-Orbit Depth, Radius, Diffuseness
0.850                   Neutron Beta
71.000   1.220  0.650   Nonlocal Proton Real Volume Depth, Radius, Diffuseness
0.000    0.000  0.000   Nonlocal Proton Imag Volume Depth, Radius, Diffuseness
0.000    0.000  0.000   Nonlocal Proton Real Surface Depth, Radius, Diffuseness
15.000   1.220  0.470   Nonlocal Proton Imag Surface Depth, Radius, Diffuseness
0.000    0.000  0.000   Nonlocal Proton Real Spin-Orbit Depth, Radius, Diffuseness
0.000    0.000  0.000   Nonlocal Proton Imag Spin-Orbit Depth, Radius, Diffuseness
0.850                   Proton Beta
 
1     49                Mass number of projectile and target (NUCLEON SCATTERING STATE)
1     20                Charge of projectile and target
0.5   1.5               Spin of projectile and target
1.0   1.0               Parity of projectile and target
1                       What Local Potential To Use? (1=User defined, 2=Pre-defined)
47.842   1.251  0.629   Local Real Volume Depth, Radius, Diffuseness
0.000    0.000  0.000   Local Imag Volume Depth, Radius, Diffuseness
0.000    0.000  0.000   Local Real Surface Depth, Radius, Diffuseness
8.697    1.236  0.440   Local Imag Surface Depth, Radius, Diffuseness
7.180    1.220  0.650   Local Real Spin-Orbit Depth, Radius, Diffuseness
0.000    0.000  0.000   Local Imag Spin-Orbit Depth, Radius, Diffuseness
1.220                   Local Coulomb Radius
1                       Use Nonlocality? [0]=No, [1]=Yes
1                       What Nonlocal Potential? (1=User defined, 2=Pre-defined, 3=Read in)
0.000    0.000  0.000   Local Part of NL: Real Volume Depth, Radius, Diffuseness
0.000    0.000  0.000   Local Part of NL: Imag Volume Depth, Radius, Diffuseness
0.000    0.000  0.000   Local Part of NL: Real Surface Depth, Radius, Diffuseness
0.000    0.000  0.000   Local Part of NL: Imag Surface Depth, Radius, Diffuseness
7.180    1.220  0.650   Local Part of NL: Real Spin-Orbit Depth, Radius, Diffuseness
0.000    0.000  0.000   Local Part of NL: Imag Spin-Orbit Depth, Radius, Diffuseness
1.220                   Local Part of NL: Coulomb Radius
71.000   1.220  0.650   Nonlocal Real Volume Depth, Radius, Diffuseness
0.000    0.000  0.000   Nonlocal Imag Volume Depth, Radius, Diffuseness
0.000    0.000  0.000   Nonlocal Real Surface Depth, Radius, Diffuseness
15.000   1.220  0.470   Nonlocal Imag Surface Depth, Radius, Diffuseness
0.000    0.000  0.000   Nonlocal Real Spin-Orbit Depth, Radius, Diffuseness
0.000    0.000  0.000   Nonlocal Imag Spin-Orbit Depth, Radius, Diffuseness
0.850                   Beta: Range of Nonlocality

931.4940                Mass Unit
20                      Number of Mesh Points for Gaussian NL Integration
0.00001                 Step in theta when calculating the spherical harmonics
2.00000                 Rmatch for n+p Bound State
2.50000                 Matching radius for nucleon bound state
-20.00000               Energy to start searching for bound state
0.00100                 Energy step when scanning for bound state
20.00000                Energy to back up after each iteration in nonlocal bound state
0.00100                 Percent diff in energy at convergence of nonlocal bound state
0.00100                 % Diff of Log Deriv. at Convergence
0.05000                 Step size for the source in d+A nonlocal integration
15.00000                Maximum radius to calculate nonlocal adiabatic source
4.00000                 Maximum radius of dr integral in nonlocal adiabatic source
1.20000                 Maximum radius of ds integral in nonlocal adiabatic source
15                      Number Of Mesh Points for d+A Source, dr Integral
2                       Number Of Mesh Points for d+A Source, theta_r Integral
30                      Number Of Mesh Points for d+A Source, ds Integral
30                      Number Of Mesh Points for d+A Source, theta_s Integral
6                       Number Of Mesh Points for d+A Source, phi_s Integral
20.00000                Maximum value of dR integral in calculation of T-matrix
20.00000                Maximum value of dr integral in calculation of T-matrix
30                      Number Of Mesh Points for T-Matrix Angular Integral
30                      Number Of Mesh Points for T-Matrix dR Radial Integral
30                      Number Of Mesh Points for T-Matrix dr Radial Integral
1.00000                 Angular step when calculating transfer cross section
1.00000                 Angular step when calculating elastic cross section

dp48Ca_20-0_NL
1                       DeuteronBoundWF.txt (0=DO NOT PRINT, 1=PRINT)
1                       NucleonBoundWF.txt
1                       DeuteronScatWFs.txt
1                       NucleonScatWFs.txt
1                       LocalBoundWF.txt
1                       NonlocalBoundWF.txt
1                       DeuteronLocalIntegral.txt
1                       NucleonLocalIntegral.txt
1                       DeuteronNonlocalIntegral.txt
1                       NucleonNonlocalIntegral.txt
1                       DeuteronLocalSmatrix.txt
1                       NucleonLocalSmatrix.txt
1                       DeuteronNonlocalSmatrix.txt
1                       NucleonNonlocalSmatrix.txt
1                       DeuteronRatioToRuth.txt
1                       NucleonRatioToRuth.txt
1                       DeuteronElasticCS.txt
1                       NucleonElasticCS.txt
1                       TransferCS.txt
\end{Verbatim} }

In the example provided before, all of the potentials were defined by the user. 
The local and nonlocal binding potentials are chosen to reproduce the neutron binding energy.
The nonlocal scattering potential used in this example is the Perey-Buck potential.
As an alternative, we can use pre-defined local and nonlocal scattering potentials to describe the same reaction.
An example of such an input file is given below.

\IB{ 
\begin{Verbatim}
5                       What Calculation? 1=n+p (B), 2=N+A (B), 3=d+A (S), 4=N+A (S), 5=(d,N), 6=(N,d)
 
0.010                   Step Size
30.000                  Maximum Radius
20.000                  Elab
20                      Lmax
2.946                   Q-Value
 
1                       Pre-Defined Deuteron Binding Potential. [1]=Gaussian
 
1     48                Mass number of fragement and core (N+A BOUND STATE)
0     20                Charge of fragement and core
0.5   0.0               Spin of fragement and core
1.0   1.0               Parity of fragement and core
1     1.5               L and J of Bound State
2                       Number of nodes in bound state?
1                       What Local Potential To Use? (1=User defined)
50.301   1.250  0.650   Local Real Volume Depth, Radius, Diffuseness
0.000    0.000  0.000   Local Real Surface Depth, Radius, Diffuseness
6.000    1.250  0.650   Local Real Spin-Orbit Depth, Radius, Diffuseness
1.250                   Local Coulomb Radius
1                       Use Nonlocality? [0]=No, [1]=Yes
1                       What Nonlocal Potential? (1=User defined, 2=Read in)
0.000    0.000  0.000   Local Part of NL: Real Volume Depth, Radius, Diffuseness
0.000    0.000  0.000   Local Part of NL: Real Surface Depth, Radius, Diffuseness
6.000    1.250  0.650   Local Part of NL: Real Spin-Orbit Depth, Radius, Diffuseness
1.220                   Local Part of NL: Coulomb Radius
63.040   1.250  0.650   Nonlocal Real Volume Depth, Radius, Diffuseness
0.000    0.000  0.000   Nonlocal Real Surface Depth, Radius, Diffuseness
0.000    0.000  0.000   Nonlocal Real Spin-Orbit Depth, Radius, Diffuseness
0.850                   Beta: Range of Nonlocality
 
2     48                Mass number of projectile and target (NUCLEON SCATTERING STATE)
1     20                Charge of projectile and target
1.0   0.0               Spin of projectile and target
1.0   1.0               Parity of projectile and target
2                       What Local Potential? (1=Deuteron Optical Potential, 2=Nucleon Potentials)
2                       What Local Nucleon Potentials To Use? (1=User defined, 2=Pre-defined)
1                       What Pre-Defined Local Nucleon Potentials for the Deuteron? (1=KD, 2=CH89)
1                       Use Nonlocality? [0]=No, [1]=Yes
2                       What Type of Nonlocal Potential? (1=Deuteron Potential, 2=Nucleon Potentials)
2                       What Local Nucleon Potentials To Use? (1=User defined, 2=Pre-defined)
1                       What Pre-Defined Local Potential? (1=Perey-Buck, 2=TPM)
 
1     49                Mass number of projectile and target (NUCLEON SCATTERING STATE)
1     20                Charge of projectile and target
0.5   1.5               Spin of projectile and target
1.0   1.0               Parity of projectile and target
2                       What Local Potential To Use? (1=User defined, 2=Pre-defined)
1                       What Pre-Defined Local Potential? (1=KD, 2=CH89)
1                       Use Nonlocality? [0]=No, [1]=Yes
2                       What Nonlocal Potential? (1=User defined, 2=Pre-defined, 3=Read in)
1                       What Pre-Defined Nonlocal Potential? (1=Perey-Buck, 2=TPM)
 
931.4940                Mass Unit
20                      Number of Mesh Points for Gaussian NL Integration
0.00001                 Step in theta when calculating the spherical harmonics
2.00000                 Rmatch for n+p Bound State
2.50000                 Matching radius for nucleon bound state
-20.00000               Energy to start searching for bound state
0.00100                 Energy step when scanning for bound state
20.00000                Energy to back up after each iteration in nonlocal bound state
0.00100                 Percent diff in energy at convergence of nonlocal bound state
0.00100                 % Diff of Log Deriv. at Convergence
0.05000                 Step size for the source in d+A nonlocal integration
15.00000                Maximum radius to calculate nonlocal adiabatic source
4.00000                 Maximum radius of dr integral in nonlocal adiabatic source
1.20000                 Maximum radius of ds integral in nonlocal adiabatic source
15                      Number Of Mesh Points for d+A Source, dr Integral
2                       Number Of Mesh Points for d+A Source, theta_r Integral
30                      Number Of Mesh Points for d+A Source, ds Integral
30                      Number Of Mesh Points for d+A Source, theta_s Integral
6                       Number Of Mesh Points for d+A Source, phi_s Integral
20.00000                Maximum value of dR integral in calculation of T-matrix
20.00000                Maximum value of dr integral in calculation of T-matrix
30                      Number Of Mesh Points for T-Matrix Angular Integral
30                      Number Of Mesh Points for T-Matrix dR Radial Integral
30                      Number Of Mesh Points for T-Matrix dr Radial Integral
1.00000                 Angular step when calculating transfer cross section
1.00000                 Angular step when calculating elastic cross section
 
dp48Ca_20-0_NL_PB
1                       DeuteronBoundWF.txt (0=Do not print, 1=Print)
1                       NucleonBoundWF.txt
1                       DeuteronScatWFs.txt
1                       NucleonScatWFs.txt
1                       LocalBoundWF.txt
1                       NonlocalBoundWF.txt
1                       DeuteronLocalIntegral.txt
1                       NucleonLocalIntegral.txt
1                       DeuteronNonlocalIntegral.txt
1                       NucleonNonlocalIntegral.txt
1                       DeuteronLocalSmatrix.txt
1                       NucleonLocalSmatrix.txt
1                       DeuteronNonlocalSmatrix.txt
1                       NucleonNonlocalSmatrix.txt
1                       DeuteronRatioToRuth.txt
1                       NucleonRatioToRuth.txt
1                       DeuteronElasticCS.txt
1                       NucleonElasticCS.txt
1                       TransferCS.txt
\end{Verbatim} }

\subsection{Compiling and running}
\label{sec-compile}

The code package contains $26$ subroutines written in Fortran 90. After downloading the source code \textit{NLAT.tar.gz}, one should unzip the tar file:

\begin{verbatim}
gunzip NLAT.tar.gz
tar -xvf NLAT.tar
\end{verbatim}

\noindent This will create the directory \textbf{NLAT}, which is organized as follows

\begin{itemize}
\item{\textbf{SOURCE/}} contains all the source code files 
\item{\textbf{makefile\_ifort}} make file for the ifort compiler 
\item{\textbf{makefile\_gfortran}} make file for the gfortran compiler 
\item{\textbf{make\_input.f90}} code to make an input file 
\item{\textbf{LOCAL\_SAMPLE/}} contains sample input and output files for a local calculation
\item{\textbf{NONLOCAL\_SAMPLE/}} contains sample input and output files for a nonlocal calculation
\end{itemize}

\noindent copy the necessary makefile into the \textbf{SOURCE} directory, renaming it to \textit{makefile}. Move to the \textbf{SOURCE} directory, and type:

\begin{verbatim}
make install clean
\end{verbatim}

\noindent This will make the executable \textit{NLAT}, which will be placed in the directory containing the \textbf{SOURCE} directory. For the input file generator, compile using the ifort compiler by typing

\begin{verbatim}
ifort -o make-input make_input.f90
\end{verbatim}

\noindent or with gfortran compiler

\begin{verbatim}
gfortran -o make-input make_input.f90
\end{verbatim}

\noindent to generate the executable for the input file maker.

In order to run the code, one first runs \textit{make-input} and is guided through a set of questions to create the input file that is stored in \textit{inputfile.in}.
Subsequently, one runs the code:
\begin{verbatim}
./NLAT < inputfile.in > output
\end{verbatim}
to produce the desired outputs.

\subsection{Output}

In addition to the default output print to screen, which contain basic properties of the bound and scattering states, as well as the transfer differential cross sections, after each run a series of output files are generated. When making the input file, the user has the option to either print all or none of the files listed below. By editing the input file, the user can select the desired output files from the provided list. Note that if all output files are generated, this may require up to 100 GB of disk space. 

\begin{itemize}
  \item{\textit{DeuteronBoundWF.txt}: The deuteron bound state wave function.} 
  \item{\textit{NucleonBoundWF.txt}:} The nucleon bound state wave function.
  \item{\textit{DeuteronScatWFs.txt}: The deuteron scattering wave function for each partial. File contains the radius, real part of wave function, and imaginary part of wave function. }
  \item{\textit{NucleonScatWFs.txt}: The nucleon scattering wave function for each partial wave. File contains the radius, real part of wave function, and imaginary part of wave function.}
  \item{\textit{LocalBoundWF.txt}: The bound state wave function resulting from the local potential.}
  \item{\textit{NonlocalBoundWF.txt}: The bound state wave function resulting from the nonlocal potential.}
  \item{\textit{DeuteronLocalIntegral.txt}: The product of the local deuteron nuclear potential and the wave function resulting from the local deuteron scattering potential. File contains the radius, real part of this term, and imaginary part of this term. }
  \item{\textit{NucleonLocalIntegral.txt}:  The product of the local nucleon nuclear potential and the wave function resulting from the local nucleon scattering potential. File contains the radius, real part of this term, and imaginary part of this term.}
  \item{\textit{DeuteronNonlocalIntegral.txt}: Just the integral term in the r.h.s. of Eq. (\ref{NLeqn3}) in the deuteron calculation at convergence. File contains the radius, real part of this term, and imaginary part of this term.}
  \item{\textit{NucleonNonlocalIntegral.txt}: Just the integral term in the r.h.s. of Eq. (\ref{NLeqn3}) in the nucleon calculation at convergence. File contains the radius, real part of this term, and imaginary part of this term.}
  \item{\textit{DeuteronLocalSmatrix.txt}:  The S-matrix elements in the deuteron calculation when using the local scattering potential.}
  \item{\textit{NucleonLocalSmatrix.txt}: The S-matrix elements in the nucleon calculation when using the local scattering potential. }
  \item{\textit{DeuteronNonlocalSmatrix.txt}: The S-matrix elements in the deuteron calculation when using the nonlocal scattering potential. }
  \item{\textit{NucleonNonlocalSmatrix.txt}:  The S-matrix elements in the nucleon calculation when using the nonlocal scattering potential. }
  \item{\textit{DeuteronRatioToRuth.txt}: Deuteron elastic differential cross section normalized to Rutherford. }
  \item{\textit{NucleonRatioToRuth.txt}: Nucleon elastic differential cross section normalized to Rutherford.  }
  \item{\textit{DeuteronElasticCS.txt}: Elastic differential cross section for the deuteron in mb/sr.  }
  \item{\textit{NucleonElasticCS.txt}: Elastic differential cross section for the nucleon in mb/sr.  }
  \item{\textit{TransferCS.txt}: Transfer cross section in mb/sr. }
\end{itemize}


\section{Summary and Conclusions}
\label{sec:sum}

Deuteron induced transfer reactions are one of the most popular tools to study single-particle structure in nuclei. Over the last couple of decades it has been established that deuteron breakup needs to be carefully taken into account in the description of (d,p) reactions. Recent studies have now demonstrated that optical potentials including nonlocality can hold very different results when compared to their local phase equivalent potentials \cite{Titus_prc2014,Ross_prc2015,Titus_prc2015}. The approximate method of including nonlocality in these calculations (the so-called Perey correction) has also been shown to be inaccurate \cite{Titus_prc2014}. 

We have implemented a suite of codes named NLAT, which perform calculations of transfer cross sections for single-nucleon transfer reactions of the type (d,p), (p,d), (d,n) and (n,d). Our implementation allows for cross sections to be calculated within the distorted wave Born approximation or within the adiabatic wave approximation. In either case, the new element of NLAT, as compared to others in the field, is that it allows for the explicit inclusion of nonlocality in the optical potentials. NLAT is general in the iterative method it uses for solving the integro-differential equations, therefore any form of nonlocality can be introduced. 

In this paper, we have summarized the reaction theory necessary to understand our implementation and briefly described the methods used to solving the intregro-differential equations. We have provided a number of checks of the various elements of NLAT and a detailed explanation of the inputs and outputs. 
We hope NLAT can provide an upgrade for the wider community interested in exploring transfer  as a means to study nuclear structure and/or analyzing transfer reaction data in our field.


\section*{Acknowledgments}

\noindent We are grateful to Ian Thompson and Gregory Potel for their help in testing the code.
This work was supported by the National Science Foundation under Grants No. PHY-1068571 and PHY-1403906 and the
Department of Energy under Contract No. DE-FG52-08NA28552.

\bibliography{cpc-nl}
\bibliographystyle{elsarticle-num}


\end{document}